\documentclass[preprint,showpacs,preprintnumbers,amsmath,amssymb,groupedaddress]{revtex4}

\usepackage{graphicx}
\usepackage{dcolumn}
\usepackage{bm}
\usepackage[usenames]{color}

\newcommand{\w}{\omega}
\newcommand{\kp}{k_{\parallel}}

\newcommand{\kpv}{\mathbf{k}_{\parallel}}

\newcommand{\eps}{\varepsilon}
\newcommand{\epsk}{\hat{\varepsilon}_{\kpv}}
\newcommand{\epsd}{\hat{\eps}_{\rm d}}

\newcommand{\wg}{\omega_{\kpv\alpha}}
\newcommand{\wo}{\omega_{0}}
\newcommand{\ud}{\mathrm{d}}
\newcommand{\rhob}{\boldsymbol{\rho}}
\newcommand{\ad}{\hat{a}_{\mathbf{k}n}}
\newcommand{\ac}{\hat{a}_{\mathbf{k}n}^{\dag}}
\newcommand{\md}{|\mathbf{d}|}
\begin{document}

\preprint{APS/123-QED}

\title{Quantum theory of spontaneous emission
in multilayer dielectric structures}
\author{Celestino Creatore}
\affiliation{Department of Physics, Politecnico di Torino, C.so Duca degli Abruzzi 24, 10129 Torino, Italy}
\affiliation{Department of Physics ``A. Volta'', Universit$\grave{a}$ degli Studi di Pavia,
via Bassi 6, I-27100, Pavia, Italy}
\author{Lucio Claudio Andreani}
\affiliation{Department of Physics ``A. Volta'', Universit$\grave{a}$ degli Studi di Pavia,
via Bassi 6, I-27100, Pavia, Italy}
\date{\today}

\begin{abstract}
We present a fully quantum-electrodynamical formalism suitable to
evaluate the spontaneous emission rate and pattern from a dipole embedded in a
non-absorbing and lossless multilayer dielectric structure. In the
model here developed the electromagnetic field is quantized by a
proper choice of a complete and orthonormal set of classical spatial
modes, which consists of guided and radiative (partially and fully)
states. In particular, by choosing a set of radiative states
characterized by a single outgoing component, we get rid of the
problem related to the quantum interference between different outgoing modes, which arises when
the standard radiative basis is used to calculate spontaneous emission patterns. After the derivation of the
local density of states, the analytical expressions for the emission
rates are obtained within the framework of perturbation theory.
First we apply our model to realistic Silicon-based structures such
as a single Silicon/air interface and a Silicon waveguide in both
the symmetric and asymmetric configurations. Then, we focus on the
analysis of the spontaneous emission process in a
silicon-on-insulator (SOI) Slot waveguide (a 6 layers model
structure) doped with Er$^{3+}$ ions (emitting at the telecom
wavelength). In this latter case we find a very good agreement with
the experimental evidence [M. Galli \textit{et al.}, Appl. Phys.
Lett. \textbf{89}, 241114 (2006)] of an enhanced TM/TE
photoluminescence signal. Hence, this model is relevant to study the
spontaneous emission in Silicon-based multilayer structures which
nowadays play a fundamental role for the development of highly
integrated multifunctional devices.
\end{abstract}

\maketitle
\section{Introduction}
It is well known that the environment surrounding an excited atom
affects its rate of spontaneous emission (SE): enhanced SE in a
resonant cavity was first predicted in the pioneering work by
Purcell~\cite{Purcell} and, later on, an inhibited SE in a small
cavity was shown by Kleppner~\cite{Kleppner81}. Such an effect can
be explained either by classical electromagnetism, in terms of a
self-driven dipole due to the reflected field at the dipole
position, or in the framework of quantum electrodynamics, as
emission stimulated by zero-point fluctuations of the
electromagnetic field. As long as the coupling between the atom and
the field is weak, both descriptions 
yield the same results~\cite{Hinds94}. In such a weak coupling
regime, the SE rate can be calculated within first order
perturbation theory by applying the Fermi's Golden Rule, and is
proportional to the local coupling of the atomic dipole moment to
the allowed photon modes, i.e. to the local density of states (LDoS)
~\cite{VanTiggelen94,VanTiggelen96}. The modifications in the
electromagnetic boundary conditions induced by the surrounding
material alter the density of states as well as the SE rate: when
the LDoS vanishes, then the SE process is inhibited, while, when an
increase in the density of states occurs, the rate of SE can be
enhanced over the free space value.

A large amount of work, both theoretical and experimental, has been
devoted to the analysis of the SE from emitters (such as atoms,
molecules or electron-hole pairs) embedded in dielectric
environments of varying complexity. In an homogeneous medium with
dielectric constant $\eps$, it has been shown by
Glauber~\cite{Glauber91} that the SE rate relative to the free space
value, is enhanced when $\eps>1$ and reduced for $\eps<1$, as it has
been demonstrated also experimentally by
Yablonovitch~\cite{Yablonovich88}. In spite of, or rather, thanks to
its relative simplicity, the single interface has been subject of a
constant
research~\cite{Drexage74,Morawitz74,Wylie84,Wylie85,Loudon91,Zakowicz94,Polman95,Barnes98,Inoue01,Wang05}
which still goes on, since it is the ideal system where experimental
and theoretical analysis can be performed in order to get the basic
understanding necessary to investigate more complex structures. The
double interface has also been widely studied, especially as a
waveguide slab, i.e., an high-index core surrounded by low-index
cladding layers, with both a quantum electrodynamical
~\cite{Loudon92,Zakowicz95b,Nha96,Urbach98,Ho99} and
classical~\cite{Brueck00} approach. In systems characterized by more
than two interfaces, both the technology and the theory needed
become more demanding, but the expected effects turn out to be also
more interesting. For instance, among multiple dielectric layers
structures, planar microcavities have been subject of intense
research in last
years~\cite{Bjork91,Rigneault96,Rigneault97,Benisty98}, due to their
ability to considerably affect the density of states and thus
strongly modify the emission into a particular mode, which is of
crucial importance for the development of light emitting devices.

In this work we study the SE rate in a non-dispersive and lossless
multilayer dielectric structure by applying a fully quantum
electrodynamical formalism. With respect to previous published
works, which generally deal with a specific dielectric structure,
our main aim is to develop a model suitable for more than one
configuration, thus taking into account all the possible modes (and
the related SE rates) which can be excited in the examined
structure. While our discussion tackles the problem of the
spontaneous emission from a theoretical point of view, the results
derived can provide a useful quantitative insight into the
modifications of the atomic radiative processes which occur in
realistic structures. As an example, we apply our method to evaluate
the SE rate in silicon-based optical waveguides, which nowadays can
be tailored according to different geometries, from simple
waveguides (three-layers geometry) to multilayers-like
configurations. These structures are characterized by a high index
contrast and are able to confine and guide the light very
efficiently in nanometer-size spatial regions as a result of total
internal reflection. The waveguiding and confining properties,
together with the low propagation losses (typical of Silicon) and
the good compatibility with complementary metal oxide semiconductor
technology, make them very attractive for the future development of
highly integrated multifunctional optoelectronic and photonic devices
(see e.g.,
Refs.~\cite{Lee01,Vlasov04,Lipson04N}). Furthermore, with an embedded
optically active layer, these structures can also be exploited as
silicon-based optical sources. With this purpose, one of the most
promising configurations recently proposed is the Slot waveguide
~\cite{Lipson04a,Lipson04b}: this system consists of a thin
(few nanometers thick) layer (slot) of low-index material bounded
by two high-index material regions (typically Silicon),
which are the core of an optical waveguide; the high-index contrast
interfaces at the slot are able to concentrate the electromagnetic
field in very narrow spatial regions ($\ll\lambda$), thus leading to
an enhancement in the radiation-matter interaction. A theoretical
investigation~\cite{Lipson05} of the emission properties of a slot
waveguide doped with Erbium ions and embedded in optoelectronic
devices, as well as the experimental evidence~\cite{Galli06} of an
enhanced light-matter coupling, have been already presented, but a
full quantum mechanical analysis of the spontaneous emission
processes in this kind of structures is still lacking. Here, we
face this problem by applying the developed formalism to
evaluate the SE rate of a dipole embedded first in a single
Silicon/air interface, then in both a symmetric (high-index
contrast) and an asymmetric Silicon waveguide, and finally in a Slot
waveguide.

In order to build up a quantum electrodynamical theory of the SE
process, the electromagnetic field must be first decomposed into the
normal modes supported by dielectric structure under consideration.
This is needed in order to set up a second quantized form of the
electromagnetic field, and then to express the local density of
states and SE rate by application of Fermi's Golden Rule. 
The LDoS can be also derived within a quantum 
electrodynamic and Green's function formalism as often done in the 
literature, using either a scalar or a dyadic Green's function, see e.g. Refs.~[\onlinecite{Wylie84,Wylie85,Hooijer01,Joulain03}]. 
In a generic multilayer structure, the normal set of modes, i.e., a
complete and orthonormal set of solutions of Maxwell equations for
the considered structure, is well known~\cite{Yeh_book} and consists
of a continuous spectrum of radiative modes and a discrete one
composed of guided modes, defined for both transverse electric (TE)
and transverse magnetic (TM) polarizations~\cite{note_radiative}. 
Guided modes are trapped
by the highest refractive index layer (if any), and are evanescent
in both half spaces - the lower and upper cladding - surrounding the
multilayer structure. Radiative modes can be either fully or
partially radiative. The former, similar to free space modes, extend
over the whole space and propagate out of the dielectric structure
from both cladding layers as outgoing plane waves, while partially
radiative modes propagate from the cladding layer with higher
refractive index only, being evanescent (due to total internal
reflection) along the lower refractive index cladding. The modes,
found as the elementary solutions of Maxwell equations with proper
boundary conditions, have more than one representation, since one
needs to characterize the asymptotic behavior of the radiative
states, such a characterization being not unique. The standard set
of radiative modes, originally introduced by Carniglia and
Mandel~\cite{Carniglia71}, and which is generally applied to
describe the interaction of a radiating system with the
electromagnetic field in a dielectric structure, is not very
convenient for SE analysis though. In this paper, we chose to apply
a set of radiative modes characterized by a single outgoing
component only. Such a choice leads to a simple definition of the LDoS for radiative states, avoiding
the difficulties related to the treatment of the interference
between different outgoing modes (see Refs.~[\onlinecite{Zakowicz95a,Glauber95}]), which arise
when the standard set of radiative modes based on the triplet
incident-reflected-transmitted waves is used. 
Furthermore, the emission rates in the lower and upper half-spaces of a generic multilayer structures 
- or, in general, the SE patterns - can be easily calculated.

The paper is organized as follows. In Sec.~II the field modes supported by
multilayer dielectric structure are listed and described. We show that the basis
of radiative states which has been used for the quantization of the electromagnetic
field in the considered dielectric structure, can be obtained from the standard set of radiative
modes by a Time-Reversal transformation. In Sec.~III we perform a standard quantization
of the electromagnetic field, and in Sec.~IV a
second quantized form for the atom-field interaction term of the whole system
Hamiltonian is set up and then used (in the electric dipole approximation)
to derive the expressions of the LDoS and the SE rate as a function of the dipole position.
In Sec.~V the spatial dependence of the SE rate will be examined for several structures of interest.
A short summary of the results is given in Sec.~VI.

\section{\label{sec:modes}System and Field Modes}
\begin{center}
\begin{figure}[ht!]
  \includegraphics*[width=0.30\textwidth]{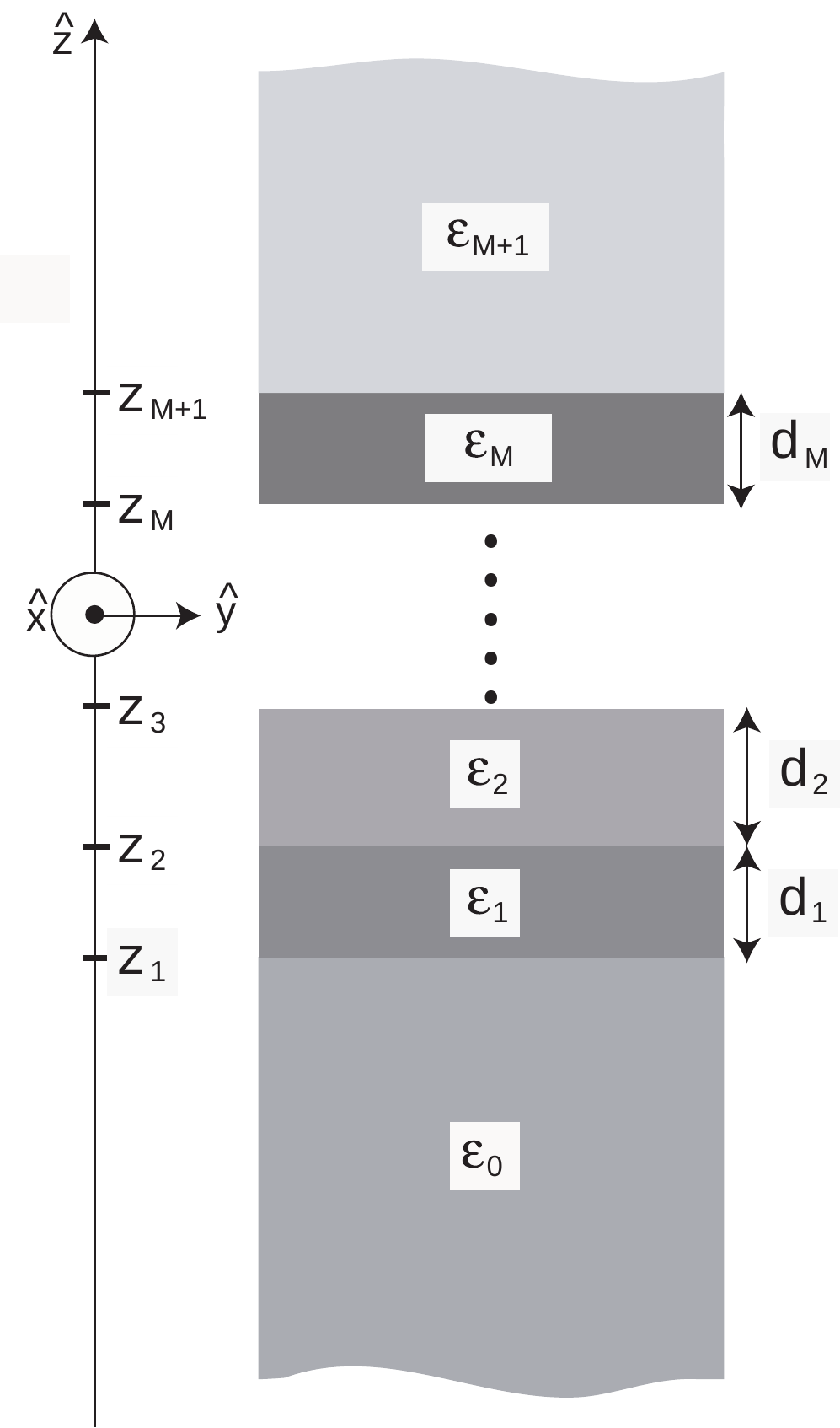}\\
  \caption{Schematic view of the multilayer dielectric structure.
  The lower (layer 0) and the upper (layer $\rm M+1$) claddings with
  dielectric constants $\eps_{0}$ and $\eps_{\rm M+1}$,
  respectively,
  are taken to be semi-infinite and surround
  the stack made by M dielectric layers, each one having
  a thickness of $d_j$ and characterized by an average dielectric
  constant $\eps_{j}$, $j=1,...,\rm M$.}\label{fig:structure}
\end{figure}
\end {center}

The system we are investigating is depicted in
Fig.~\ref{fig:structure}: it is made up of M dielectric layers (stack) which
are parallel to the $xy$ plane and assumed to be infinite along the
$x$ and $y$ directions. Each layer is $d_j$ ($j=1,...,\rm M$) thick
and the surrounding media, i.e., the lower (layer 0) and the upper
(layer M+1) claddings, are taken to be semi-infinite. Each of the
$\rm M+2$ media is supposed to be lossless, isotropic, and
homogeneous along the vertical ($z$) direction. Hence, the
dielectric constant $\eps(\mathbf{r})=\eps(\rhob,z)$ is a piecewise
constant function in the $z$ direction and it will be denoted as
$\eps_{j}=\eps_{j}(z)$ in each of the $\rm M+2$ dielectric
media.
In order to develop a quantum theory for the spontaneous emission of
a dipole embedded in such a dielectric structure, the classical
electromagnetic modes, which are needed in the expansion of the
electromagnetic field operators (see Sec.~\ref{sec:quantization}),
must be first specified. The modes are found as the solutions
of the following eigenvalue problem
\begin{equation}\label{eq:eigen-H}
    \boldsymbol{\nabla}\times\left[\frac{1}{\eps(\mathbf{r})}
    \boldsymbol{\nabla}\times\mathbf{H}\right]=\frac{\w^2}{c^2}\mathbf{H}\,,
\end{equation}
which results from the homogenous Maxwell equations for the electric
and magnetic fields $\mathbf{E}$, $\mathbf{H}$ having harmonic time
dependence $\rm exp(-i\w t)$, and with the condition
$\boldsymbol{\nabla}\cdot\mathbf{H}=0$ being fulfilled. The set of
these fundamental modes is complete and orthonormal,
\begin{equation}
    \mathbf{H}(\mathbf{r})=\sum_{\mu}c_{\mu}\mathbf{H}_{\mu}(\mathbf{r})\,,
\end{equation}
the orthonormality condition being expressed by
\begin{equation}\label{eq:ortho-H}
    \int\mathbf{H}^{*}_{\mu}(\mathbf{r})\cdot
    \mathbf{H}_{\nu}(\mathbf{r})\ud\mathbf{r}=\delta_{\mu\nu}\,.
\end{equation}
The electric field eigenmodes, which can be obtained from
$\displaystyle \mathbf{E}(\mathbf{r})=i\frac{c}{\w\eps(\mathbf{r})}
\boldsymbol{\nabla}\times\mathbf{H}(\mathbf{r})$, are also orthonormal
according to the following condition~\cite{Carniglia71}:
\begin{equation}
    \int\eps(\mathbf{r})\mathbf{E}^{*}_{\mu}(\mathbf{r})\cdot
    \mathbf{E}_{\nu}(\mathbf{r})\ud\mathbf{r}=\delta_{\mu\nu}\,.
\end{equation}

Since the whole system is homogenous in the $xy$ plane the field
modes will be factorized as $\displaystyle
\mathbf{E}(\mathbf{r},t)\,[\mathbf{H}(\mathbf{r},t)]\,=e^{-i\w t +
i\kpv\cdot\rhob}\,\mathbf{E}(z)\,[\mathbf{H}(z)]$, where
$\kpv=\kp\hat{k}_{\parallel}=(k_{x},k_{y})$ is the in-plane
propagation vector.

In a lossless multilayer dielectric structure, the complete set of
orthonormal modes consists of an infinite number of radiative modes
and a finite number of guided modes. The former can be classified
into two types. Fully radiative modes, akin to free-space modes,
radiate in both the lower and upper cladding, while partially
radiative modes radiate only in the cladding with the higher
refractive index, propagating out of the smaller index cladding as
evanescent waves with exponentially decreasing amplitude. Guided
modes propagate along the dielectric planes only, being trapped
(confined) by the highest refractive index layer and characterized
by an evanescent field profile in both claddings.

Whereas the guided modes are completely specified by the Maxwell
equations and the proper continuity conditions across the dielectric
boundaries, the radiative modes are not, and their asymptotic
behavior at infinity (when $z\rightarrow\pm\infty$) has to be
characterized. Such a characterization, however, is not unique.
\begin{center}
\begin{figure}[ht!]
  \includegraphics*[width=0.50\textwidth]{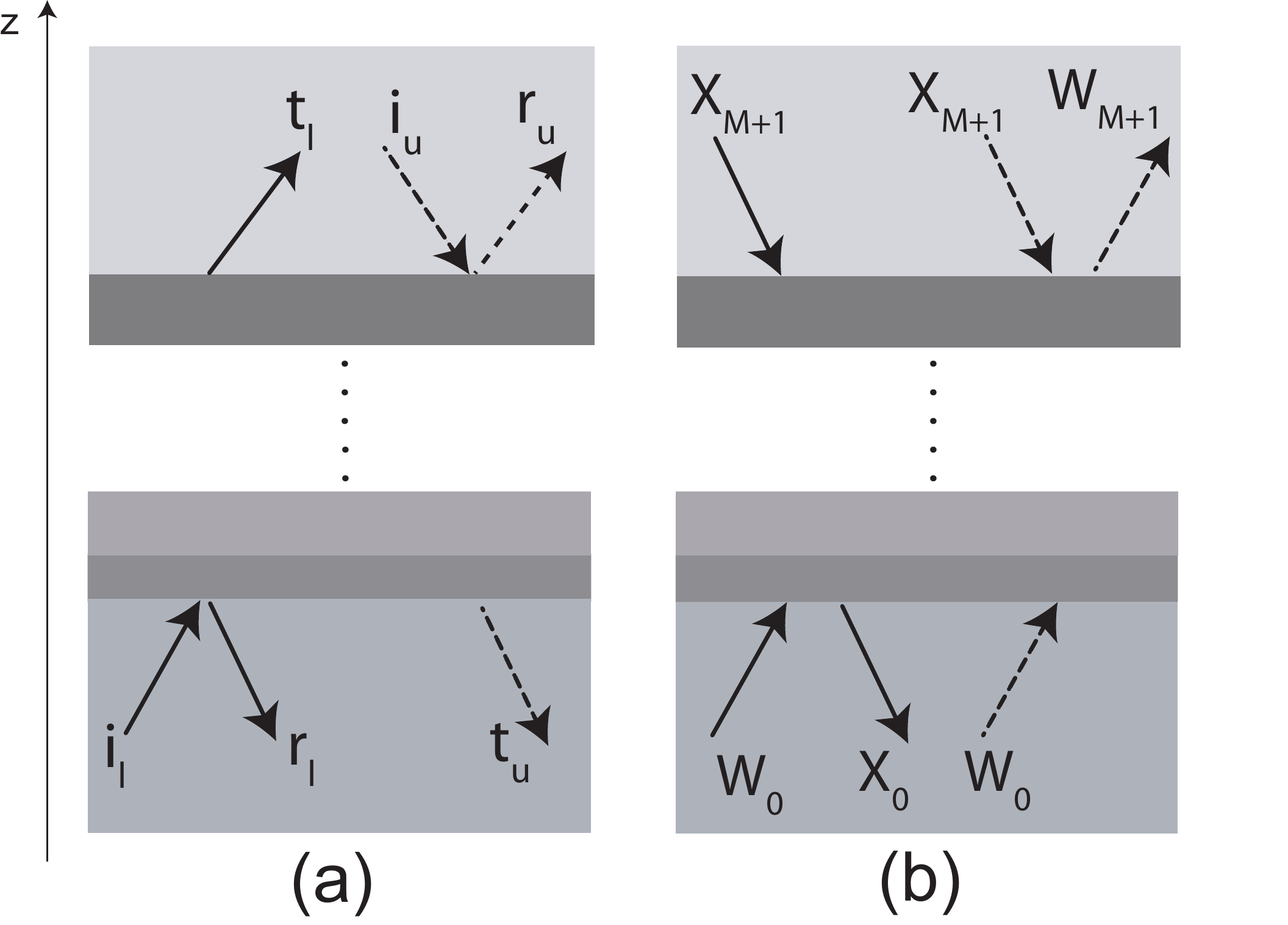}\\
  \caption{The radiative modes~\cite{note_radiative} in a multilayer dielectric
  structure. (a): the standard set of modes based on the triplets
  incident-reflected-transmitted waves, $\{\rm i_{l},\,r_{l},t_{l}\}$ for
  waves incoming from the lower cladding, $\{\rm i_{u},\,r_{u},t_{u}\}$
  for waves incoming from the upper cladding. (b): the set of
  modes specified by a single outgoing component and two incoming
  (towards the stack) waves, $\{\rm X_{0}\,,W_{0},\,X_{\rm M+1}\}$ for
  states outgoing in the lower cladding,  $\{\rm W_{\rm M+1}\,,W_{0},\,X_{\rm M+1}\}$
  for states outgoing in the upper cladding. The notation refers to
  TE-polarized modes; for TM polarization one needs the replacements
  $\rm W\rightarrow Y$ and $\rm X\rightarrow Z$.}\label{fig:rad-modes}
\end{figure}
\end{center}
The standard choice for radiative modes [see
Fig.~\ref{fig:rad-modes}(a)] assumes one incident wave incoming
towards the stack of M layers either from the lower or from the upper
cladding, and two outgoing waves, one being reflected (on the same
side of the incoming one) from the stack and the other being
transmitted (on the opposite side) across it. This set of modes,
originally introduced by Carniglia and Mandel~\cite{Carniglia71} for
the quantization of the electromagnetic field in a dielectric
interface, is orthonormal and complete~\cite{Bialinicki72} and it
has been widely employed to characterize the radiative states in
structures like dielectric waveguides~\cite{Loudon92,Urbach98} and
planar dielectric microcavities~\cite{Rigneault96,Hooijer01}.

Such a choice, however, is not the most convenient when dealing with
radiation emission analysis. As shown in
Fig.~\ref{fig:rad-modes}(a), both the reflected and the transmitted
components (the pairs $\{\rm{r}_{l},\,\rm{t}_{u}\}$ and
$\{\rm{r}_{u},\,\rm{t}_{l}\}$), which belong to two different modes,
contribute to the total emission in a given direction. As pointed out by Zakowicz \cite{Zakowicz95a}, 
the computation of the radiative density of states turns out
to be problematic since the quantum interference between the two different outgoing modes has to
be explicitly taken into account. Thus, interference terms must be 
considered when emission in either the upper or the lower layer is evaluated. 
As shown in the reply by Glauber and Lewenstein~\cite{Glauber95}, interference effects cancel out only when 
the emission in the upper cladding layer ($\{\rm{r}_{u},\,\rm{t}_{l}\}$) and in the lower one 
($\{\rm{r}_{l},\,\rm{t}_{u}\}$) are combined.  To avoid this subtle interference problem, 
and to be able to calculate the fraction of emission in either the upper or the lower cladding layer,
a more suitable way to define the radiative states in emission
problems is  to choose a set of modes based on a single
outgoing wave. This component comes together with two incoming waves propagating towards the structure, 
as shown in Fig.~\ref{fig:rad-modes}(b) with the triplets $\{\rm W_{\rm
M+1}\,,W_{0},\,X_{\rm M+1}\}$ and $\{\rm
X_{0}\,,W_{0},\,X_{\rm M+1}\}$ for states outgoing in the upper and lower cladding layers, respectively.  
By using this mode decomposition, the
total emission signal is thus completely specified by one outgoing
mode only - either by the component labeled as $\rm W_{M+1}$ for
radiative states outgoing in the upper cladding, or by the component
$\rm X_{0}$ for states outgoing in the lower cladding - 
and interference terms never arise.

It is worth to notice that this set of radiative modes can be obtained from the standard one
previously discussed (which is specified in terms of the incoming waves)
after application of Time-Reversal (TR) operator. A Time-Reversal operation transforms modes
propagating along the positive $z$-direction into modes propagating along
the negative one, and therefore a mode incoming from the upper (lower) layer,
into a mode outgoing from the upper (lower) layer.
Hence, as from Figs.~\ref{fig:rad-modes}(a)-(b), the radiative modes $\{\rm
X_{0}\,,W_{0},\,X_{\rm M+1}\}$ and $\{\rm W_{\rm
M+1}\,,W_{0},\,X_{\rm M+1}\}$ are the TR counterparts of the triplets
$\{\rm i_{l},\,r_{l},t_{l}\}$ and $\{\rm i_{u},\,r_{u},t_{u}\}$, respectively.
Furthermore, since the algebraic properties are invariant under Time-Reversal operations,
also the new set of radiative states is orthonormal and complete. The following rule for
the Time-Reversal operation over a generic spatial mode $\displaystyle \propto e^{iqz}$ propagating along the $z$-direction
with wavevector $q$, can be established:
\begin{equation}\label{eq:TR-rule}
    e^{iqz}\,\xrightarrow{\rm{TR}}\,e^{-iq^{\ast}z}\,.
\end{equation}
The transformation rule Eq.~(\ref{eq:TR-rule}) applies to both fully and partially radiative modes:
outgoing modes towards positive (negative) $z$ being $\propto e^{iqz}$ ($\propto e^{-iqz}$),
specified by the real wavevector $q$, keep the plane-wave-like character,
turning into outgoing modes towards negative (positive ) $z$ $\propto e^{-iqz}$ ($\propto e^{iqz}$).
The evanescent modes $\propto e^{-\kappa|z|}$, characterized
by the imaginary wavevector $q=i\kappa$, keep the exponentially decaying profile after the
transformation Eq.~(\ref{eq:TR-rule}).

A mode decomposition characterized by a single outgoing
component, like the one here described, has been already used to specify the radiative modes
in dielectric interfaces~\cite{Zakowicz94,Inoue01} as well as in slab waveguides~\cite{Zakowicz95b}. 
It has also been used in the formally similar problem of diffraction losses in photonic crystal waveguides~\cite{Andreani06}. 
Here, we extend its application within a quantum electrodynamical theory suitable to the analysis
of radiation emission in generic multilayer structures.

In the following a detailed description of both guided and radiative  profiles
is given.
\subsection{\label{subsec:rad_modes}Radiative modes} As previously
introduced, the set of radiative states consists of a single
outgoing component propagating outward from the whole structure and
two other waves propagating towards it. In each of the M layers the
field is a superposition of two modes propagating in opposite
directions (with respect to the $z$-direction). The modes are specified and
labeled by the propagation wavevector $\mathbf{k}=(\kpv,q)$, where
the $z$-component $q$ in each of the M+2 media, is given by
\begin{equation}\label{eq:kz-definition}
    q_{j}=\sqrt{\eps_{j}\frac{\w^2}{c^2} - \kp^2}\,,\qquad
    j=0,...,\rm M+1\,.
\end{equation}
Let us denote by $\epsk=\hat{z}\times\hat{k}_{\parallel}$ the unit
vector which is orthogonal to both $\kpv=\kp\hat{k}_{\parallel}$ and
$\hat{z}$ and set $z_{1}=-d_{1}/2$, $z_{j}=z_{j-1}+d_{j-1}$ with
$j=2,...,\rm M+1$. With the implicit time dependence $e^{-i\w t}$,
the field profiles for TE polarization are given by
\begin{equation}\label{eq:TE-rad-profs-E0}
    \mathbf{E}_{\kpv}^{\rm TE}(\rhob,z)=\frac{e^{i\kpv\cdot\rhob}}{\sqrt{V}}i\epsk
    E^{\rm TE}(\kp,z)\,,
\end{equation}
\begin{equation}\label{eq:TE-rad-profs-H0}
    \mathbf{H}_{\kpv}^{\rm TE}(\rhob,z)=\frac{e^{i\kpv\cdot\rhob}}{\sqrt{V}}i\frac{c}{\w}
    [H^{\rm TE}_{\perp}(\kp,z)\hat{z} + H^{\rm
    TE}_{\parallel}(\kp,z)\hat{k}_{\parallel}]\,,
\end{equation}
where $V$ is a normalization box-volume which disappears in the final results. 
The expressions for the amplitudes $E^{\rm TE}$, $H^{\rm
TE}_{\perp}$ and $H^{\rm TE}_{\parallel}$ as well as the method used
to obtain them are detailed in Appendix \ref{subapp:rad_modes}.

For TM-polarized radiative modes the field profiles are given by:
\begin{equation}\label{eq:TM-rad-profsH0}
     \mathbf{H}_{\kpv}^{\rm TM}(\rhob,z)=\frac{e^{i\kpv\cdot\rhob}}{\sqrt{V}}i\epsk
     H^{\rm TM}(\kp,z)\,,
\end{equation}
\begin{equation}\label{eq:TM-rad-profsE0}
    \mathbf{E}_{\kpv}^{\rm TM}(\rhob,z)=\frac{e^{i\kpv\cdot\rhob}}{\sqrt{V}}i\frac{c}{\eps_{j}\w}
    [E^{\rm TM}_{\perp}(\kp,z)\hat{z} + E^{\rm TM}_{\parallel}(\kp,z)\hat{k}_{\parallel}]\,,
\end{equation}
where $H^{\rm TM}$, $E^{\rm TM}_{\perp}$ and $E^{\rm
TM}_{\parallel}$ are given in Appendix A.

\subsection{\label{subsec:gui_modes}Guided modes}
In order for the whole dielectric structure to support a set of
guided modes, (at least) one of the dielectric constant $\eps_j$
($j=1,...,\rm M$) of the M inner layers has to fulfill the
constraint
\begin{equation}\label{eq:guided_constraint}
    \eps_{j}=\eps_{\rm max} > \eps_{0}\,,\eps_{\rm M+1}\,,
\end{equation}

The guided modes, which are in-plane propagating and evanescent
along the z direction, are labeled by the in-plane wavevector
$\kpv=\kp\hat{k}_{\parallel}$ and the mode index $\alpha$ [$\alpha$
$\geq\,1$ if Eq.~(\ref{eq:guided_constraint}) holds] in a joint
single index $\mu=(\kpv,\alpha)$. By $q_{j\,\mu}$ we denote the
$z$-component of the guided mode wavevector,
\begin{equation}
q_{j\,\mu}=\sqrt{\eps_j\frac{\w_{\mu}^2}{c^2} -
\kp^2}\,.\qquad\qquad j=1,...,\textrm{M}\,,
\end{equation}
where $\w_{\mu}=\wg$ is the frequency of the $\alpha$-th guided
mode. In the upper ($j=\rm M+1$) and lower ($j=0$) claddings
$q_{j\,\mu}$ is purely imaginary, $\displaystyle
q_{j\,\mu}=i\chi_{j\,\mu}$ where $\chi_{j\,\mu}=\sqrt{\kp^2 -
\eps_{j}\frac{\w_{\mu}^{2}}{c^2}}$, and hence the mode field
$\propto\,\textrm{exp}(\pm i q_{j\,\mu}z)$ decays exponentially
along the $z$ direction. In the following we give their explicit
form, which results from a generalization of the standard waveguide
field modes (see e.g. Refs.~\cite{Yariv,Andreani06}).

The guided modes for TE polarization are given by:
\begin{equation}\label{eq:TE-guided-profs-E0}
    \mathbf{E}_{\kpv}^{\rm TE}(\rhob,z)=\frac{e^{i\kpv\cdot\rhob}}
    {\sqrt{S}}i\frac{\w_{\mu}}{c}\epsk E^{\rm TE}(\kp,z)\,,
\end{equation}
\begin{equation}\label{eq:TE-guided-profs-H0}
    \mathbf{H}_{\kpv}^{\rm TE}(\rhob,z)=\frac{e^{i\kpv\cdot\rhob}}
    {\sqrt{S}}[H_{\perp}^{\rm TE}(\kp,z)\hat{z} +  H_{\parallel}^{\rm TE}(\kp,z)\hat{k}_{\parallel}]\,,
\end{equation}
where $S$ is a normalization surface which cancels in the final results, 
and $E^{\rm TE}$, $H_{\perp}^{\rm TE}$ and $H_{\parallel}^{\rm TE}$ 
are given in Appendix \ref{subapp:gui_modes}.
For TM polarization, the guided modes have the following field
profiles:
\begin{equation}\label{eq:TM-guided-profs-H0}
    \mathbf{H}_{\kpv}^{\rm TM}(\rhob,z)=\frac{e^{i\kpv\cdot\rhob}}
    {\sqrt{S}}\epsk H^{\rm TM}(\kp,z)\,,
\end{equation}
\begin{equation}\label{eq:TM-guided-profs-E0}
    \mathbf{E}_{\kpv}^{\rm TM}(\rhob,z)=\frac{e^{i\kpv\cdot\rhob}}{\sqrt{S}}\frac{c}{\w_{\mu}}
    [E^{\rm TM}_{\perp}(\kp,z)\hat{z} +  E^{\rm TM}_{\parallel}(\kp,z)\hat{k}_{\parallel}]\,,
\end{equation}
where $H^{\rm TM}$, $E^{\rm TM}_{\perp}$ and $E^{\rm
TM}_{\parallel}$ are given in Appendix \ref{subapp:gui_modes}.
\section{\label{sec:quantization}Field Quantization}
In this Section the canonical quantization of the electromagnetic
field in a non-uniform isotropic dielectric medium described by a
piecewise constant  permeability $\eps(\mathbf{r})$ is
performed~\cite{Glauber91,Bath06,Gerace07}. The electric
displacement vector and the magnetic induction (a unit magnetic
permeability is assumed) are simply given by the relations
\begin{equation}\label{eq:basic_rels}
\mathbf{D}=\eps(\mathbf{r})\mathbf{E}\,,\quad\mathbf{B}=\mathbf{H}\,.
\end{equation}
The starting point is the quantization of the vector potential
$\mathbf{A}$ which is defined by the familiar relations
\begin{eqnarray}\label{eq:familiar_rels}
    \mathbf{B}&=&\boldsymbol{\nabla}\times\mathbf{A}\,,\\
    \mathbf{E}&=&-\boldsymbol{\nabla}\Phi - \frac{1}{c}\frac{\partial\mathbf{A}}{\partial t}\,.
\end{eqnarray}
We use here the generalized Coulomb gauge~\cite{Glauber91} defined,
in absence of external charges, by the choice $\Phi=0$ and the relation
\begin{equation}\label{eq:gauge}
    \boldsymbol{\nabla}\cdot\left[\eps(\mathbf{r})\mathbf{A}\right]=0\,,
\end{equation}
which automatically satisfies the transversality condition on
$\mathbf{D}$,
$\boldsymbol{\nabla}\cdot\mathbf{D}=\boldsymbol{\nabla}\cdot[\eps(\mathbf{r})\dot{\mathbf{A}}]=0$,
and is consistent with the equation of motion for the vector
potential $\mathbf{A}$
\begin{equation}\label{eq:eqmotion_A}
    \boldsymbol{\nabla}\times\left(\boldsymbol{\nabla}\times\mathbf{A}\right)
    + \frac{\eps(\mathbf{r})}{c^2}\frac{\partial^{2}\mathbf{A}}{\partial\rm{t}^2}=0\,.
\end{equation}
In order to obtain a second-quantized Hamiltonian for the free
photon field, we first introduce the classical Hamiltonian function
$\mathcal{H}_{\rm em}$, i.e. the total electromagnetic energy,
\begin{eqnarray}\label{eq:Hamiltonian_classical}
    \mathcal{H}_{\rm em}=&&\frac{1}{4\pi}\int_{V}
    \boldsymbol{\Pi}(\mathbf{r},t)\dot{\mathbf{A}}(\mathbf{r},t)\ud\mathbf{r} - \mathcal{L}\nonumber\\
    &&=\frac{1}{8\pi}\int_{V}\left[\eps(\mathbf{r})\mathbf{E}(\mathbf{r})^2 + \mathbf{B}(\mathbf{r})^2\right]\ud\mathbf{r}\,,
\end{eqnarray}
where $V$ is a quantization volume,
$\boldsymbol{\Pi}=\eps(\mathbf{r})\dot{\mathbf{A}}(\mathbf{r},t)/c^{2}$
is the canonical momentum, and $\displaystyle
\mathcal{L}=\frac{1}{8\pi}\int_{V}[\eps(\mathbf{r})\mathbf{E}(\mathbf{r})^{2}
- \mathbf{B}(\mathbf{r})]\ud\mathbf{r}$ is the Lagrangian
function from which Eq.~(\ref{eq:eqmotion_A}) follows after Hamilton's principle.
The vector field operator $\hat{\mathbf{A}}$ is then
expanded in normal modes:
\begin{equation}\label{eq:A_operator}
    \displaystyle
    \hat{\mathbf{A}}=\sum_{\mathbf{k},n}(2\pi\hbar\w_{\mathbf{k}n})^{1/2}
    [\ad\mathbf{A}_{\mathbf{k}n}(\mathbf{r})e^{-i\w_{\mathbf{k}n}t} +
    \ac\mathbf{A}_{\mathbf{k}n}^{*}(\mathbf{r})e^{i\w_{\mathbf{k}n}t}]\,,
\end{equation}
where $\ac$ ($\ad$) are Bose creation (destruction) operators of
field quanta with energies $\hbar\w_{\mathbf{k}n}$ satisfying the
usual commutation relations
\begin{equation}\label{eq:Bose_comm}
    [\ad,\,\hat{a}^{\dag}_{\mathbf{k}'n'}]=\delta_{\mathbf{k},\mathbf{k}'}
    \delta_{n,n'}\,,\quad
    [\ad,\,\hat{a}_{\mathbf{k}'n'}]=[\ac,\,\hat{a}^{\dag}_{\mathbf{k}'n'}]=0\,,
\end{equation}
$n$ being a generic index labeling the corresponding
eigenmode characterized by the wavevector
$\mathbf{k}$.
\begin{equation}\label{eq:ortho-A}
    \int_{V}\eps(\mathbf{r})\mathbf{A}^{*}_{\mathbf{k}n}(\mathbf{r})\cdot
    \mathbf{A}_{\mathbf{k}'n'}(\mathbf{r})\ud\mathbf{r}=\frac{c^2}{\w^{2}_{\mathbf{k}n}}
    \delta_{\mathbf{k},\mathbf{k}'}\delta_{n,n'}\,,
\end{equation}
From Eq.~(\ref{eq:ortho-A}) the following orthonormality
conditions~\cite{Carniglia71,Glauber91,Bath06} for the electric and
magnetic fields follow:
\begin{equation}\label{eq:ortho-E}
    \int_{V}\eps(\mathbf{r})\mathbf{E}^{*}_{\mathbf{k}n}(\mathbf{r})\cdot
    \mathbf{E}_{\mathbf{k}'n'}(\mathbf{r})\ud\mathbf{r}=
    \delta_{\mathbf{k},\mathbf{k}'}\delta_{n,n'}\,,
\end{equation}
\begin{equation}\label{eq:ortho-B}
    \int_{V}\mathbf{B}^{*}_{\mathbf{k}n}(\mathbf{r})\cdot
    \mathbf{B}_{\mathbf{k}'n'}(\mathbf{r})\ud\mathbf{r}=
    \delta_{\mathbf{k},\mathbf{k}'}\delta_{n,n'}\,.
\end{equation}
Finally, from Eqs.~(\ref{eq:Hamiltonian_classical}) and
(\ref{eq:A_operator}), one gets the well known second-quantized form
for the free photon field:
\begin{equation}\label{eq:Hamiltonian-free-photons}
    \hat{H}_{\gamma}=\sum_{\mathbf{k},\,n}\hbar\w_{\mathbf{k}n}\left(\ac\ad + \frac{1}{2}\right)\,.
\end{equation}

\section{\label{sec:emission_rates}Emission Rates}
In this Section the spontaneous transition rate of an excited atom
embedded in a non-uniform dielectric medium is calculated. We
suppose that the atom, located at position  $z$ and initially in the
excited state $|\rm x\rangle$ (having energy $\hbar\w_{\rm x}$)
undergoes a spontaneous dipole transition to its ground state $|\rm
g\rangle$ (having energy $\hbar\w_{\rm g}$) thereby emitting a
photon of energy $\hbar\w_{0}=\hbar\w_{\rm x} - \hbar\w_{\rm g}$.
The total Hamiltonian of such a system can be written as
\begin{equation}\label{eq:total_Hamiltonian}
    \hat{H}=\hat{H}_{\gamma} + \hat{H}_{\rm A} + \hat{H}_{\gamma-\rm A}\,,
\end{equation}
where $\hat{H}_{\gamma}$ is the free-field Hamiltonian given by
Eq.~(\ref{eq:Hamiltonian-free-photons}), $\hat{H}_{\rm A}$ is the
free-atom Hamiltonian,
\begin{equation}\label{eq:free-atom_Hamiltonian}
    \hat{H}_{\rm A}=\hbar\w_{0}|\rm x\rangle\langle \rm x|+
    \hbar\w_{\rm g}|\rm g\rangle\langle \rm g|\,,
\end{equation}
and $\hat{H}_{\gamma-\rm A}$ is the atom-field interaction term
which, in the electric dipole approximation and near the atomic
resonance $\w\approx\w_{0}$, reads~\cite{Sakurai_book} as
\begin{equation}\label{eq:inter-Hamiltonian}
    \hat{H}_{\gamma-\rm
    A}\approx(\hat{\sigma}_{+}\mathbf{d}+\hat{\sigma}_{-}\mathbf{d}^{*})\cdot\hat{\mathbf{E}}(\mathbf{r},t)\,,
\end{equation}
where $\hat{\sigma}_{-}=|\rm g\rangle\langle \rm x|$ and
$\hat{\sigma}_{+}=|\rm x\rangle\langle \rm g|$ are the atomic down-
and atomic up- transition operators, respectively, and
$\displaystyle \mathbf{d}=\mathbf{d}_{\rm xg}=\langle\rm{
x}|\hat{\mathbf{d}}|\rm{g}\rangle=|\mathbf{d}|\hat{\eps}_{\rm d}$ is
the dipole matrix element, $\hat{\mathbf{d}}=e\hat{\mathbf{r}}$
being the atomic dipole operator of the atom located at
$\mathbf{r}$. The electric field operator
$\hat{\mathbf{E}}(\mathbf{r})$ can be obtained from the vector
potential operator $\hat{\mathbf{A}}$ through
Eqs.~(\ref{eq:familiar_rels}) and (\ref{eq:A_operator}),
\begin{eqnarray}\label{eq:E_operator}
    \displaystyle
    \hat{\mathbf{E}}(\mathbf{r},t)=&i&\sum_{\mathbf{k},n}(2\pi\hbar\w_{\mathbf{k}n})^{1/2}
    [\ad\mathbf{E}_{\mathbf{k}n}(\mathbf{r})e^{-i\w_{\mathbf{k}n}t}\nonumber\\
    &-&\ac\mathbf{E}_{\mathbf{k}n}^{*}(\mathbf{r})e^{i\w_{\mathbf{k}n}t}]\,.
\end{eqnarray}

We assume that the interaction between the excited two-level system
and the electromagnetic field in the dielectric medium is not too
strong, so that the transition between two states can be studied
within the framework of perturbation theory. Let us then consider
the initial $|i\rangle$ and the final $|f\rangle$ states of the
combined atom-radiation system: initially there are no photons and
the atom is in the upper (excited) level,
$|i\rangle=|0\rangle\otimes|\rm{x}\rangle$; in the final state one
photon is emitted in any mode of the electromagnetic field of
frequency $\w_{\mathbf{k}n}$ and the atom is in the lower (ground)
level, $|f\rangle=|1_{\mathbf{k}n}\rangle\otimes|\rm{g}\rangle$.
According to Fermi's \textit{Golden Rule} (see e.g.
Ref.~\cite{Loudon_book}) the spontaneous emission rate
$\Gamma=\Gamma(\mathbf{r})$ of an atom located at position
$\mathbf{r}$ is
\begin{equation}\label{eq:golden_rule1}
    \Gamma(\mathbf{r})=\frac{2\pi}{\hbar^2}
    \sum_{f}\left|\langle f|\hat{H}_{\gamma-\rm A}|i\rangle\right|^{2}\delta(\w_{i} - \w_{f})\,,
\end{equation}
where $\hbar\w_i$ and $\hbar\w_f$ are th energies of the initial and
final state, respectively. By insertion of Eq.~(\ref{eq:E_operator})
in the above expression, and using the commutation rules for $\ad$
and $\ac$, the spontaneous decay rate finally reads
\begin{equation}\label{eq:golden_rule2}
    \Gamma(\mathbf{r})=\frac{4\pi^2\md^{2}}{\hbar}\sum_{\mathbf{k},n}
    |\mathbf{E}_{\mathbf{k}n}(\mathbf{r})\cdot\epsd|^2
    \w_{\mathbf{k}n}\delta(\w_0 -\w_{\mathbf{k}n})\,.
\end{equation}
By taking into account the $i$-th cartesian component $E^{i}_{\mathbf{k}n}$ of the
eigenmode $\mathbf{E}_{\mathbf{k}n}(\mathbf{r})$, the contribution $\Gamma_{i}$
to the total emission rate can be written as
\begin{equation}\label{eq:golden_rule3}
    \Gamma_{i}(\mathbf{r})=\frac{4\pi^{2}\md^{2}\w_{0}}{\hbar}
    J_{i}(\w_{0},\,\mathbf{r})\,,
\end{equation}
where $J_{i}(\w_{0},\mathbf{r})$ is the $i$-th contribution to the local density of states (LDOS)
~\cite{VanTiggelen94,VanTiggelen96}
$J(\w_{0},\mathbf{r})$:
\begin{eqnarray}\label{eq:LDOS}
    J_{i}(\w_{0},\mathbf{r})&=&\sum_{n}\int\ud\mathbf{k}|E^{i}_{\mathbf{k}n}(\mathbf{r})|^{2}
    \delta(\w_0 -\w_{\mathbf{k}n})\,,\nonumber\\
    J(\w_{0},\mathbf{r})&=&\sum_{i}J_{i}(\w_{0},\mathbf{r})\,.
\end{eqnarray}

In a multilayer dielectric structure, an excited dipole can decay
either as a radiative or a guided eigenmode. As discussed in
Sec.~\ref{subsec:rad_modes}, the radiative modes are specified by
the propagation vector $(\kpv,q)$ of the outgoing component. Hence,
in Eq.~(\ref{eq:golden_rule2}), $\mathbf{k}=(\mathbf{\kpv},q)$ and
$n$=($p\,,j$) is a double index specifying the final state
parameters, namely the field polarization $p=\rm TE,\,TM$ and the
cladding layer $j$ in which the emission occurs, $j=0$ for emission
in the lower cladding and $j=\rm M+1$ for emission in the upper
cladding. For what concerns the guided modes, $\mathbf{k}=\kpv$ and
$n$=($p\,,\alpha$), where $\alpha$ is the guided mode index
introduced in Sec.~\ref{subsec:gui_modes}. Furthermore, since the
dielectric function $\eps(\mathbf{r})=\eps(z)=\eps_{j}$ is homogenous
in each layer, the spontaneous emission rate will be expressed
as a function of the $z$ coordinate only.

For both decay channels (radiative and guided)
two contributions to the total emission rate can be distinguished:
(i) the emission rate $\Gamma_{\parallel}$ due to the decay
of horizontal dipoles, i.e. in-plane oriented dipoles ($\epsd=\hat{x}$ or
$\epsd=\hat{y}$), which couple to both TE- and TM-polarized fields,
(ii) the  rate $\Gamma_{\perp}$ due to the decay
of vertical dipoles ($\epsd=\hat{z}$) which interact with
TM-polarized modes only. For randomly oriented dipoles, the total averaged
emission rate can thus be written as
$\displaystyle \Gamma=\frac{2}{3}\Gamma_{\parallel} +
\frac{1}{3}\Gamma_{\perp}$. In the rest of this Section
we derive the exact expressions for the emission rates
into both radiative and guided modes.

\subsection{Emission rates into radiative modes}
For each propagation wavevector $\mathbf{k}=(\kpv,q)$ the frequency
$\displaystyle\w_{\gamma}=\frac{c}{\sqrt{\eps_j}}(\kp^2 +
q^2)^{1/2}$ of the radiative modes has to satisfy the relation
\begin{equation}\label{eq:emission-rad-freq}
\kp^2  < k_{j}^{2}=\eps_{j}\frac{\w_{\gamma}^{2}}{c^2}\,,\qquad j=0,\,\rm M
+1\,,
\end{equation}
where $\eps_{j}=\eps_{0}$ ($\eps_{\rm M+1}$) if the emission occurs
in the lower (upper) cladding. With $\w_{\mathbf{k}n}=\w_{\gamma}$
in Eq.~(\ref{eq:golden_rule2}), the emission rate into the radiative
modes $\Gamma=\Gamma(z)$ is thus given by
\begin{eqnarray}\label{eq:Gamma_rad}
    \Gamma(z)&=&\frac{4\pi^{2}\md^2}{\hbar}\sum_{p=\rm TE,\,TM}
    \sum_{j=\rm 0,M+1}\sum_{\kpv}\sum_{q}|\mathbf{E}^{p}_{\kpv}
    (\rhob,z)\cdot\epsd|^{2}\nonumber\\
    &\times&\w_{\gamma}\delta(\omega_{0}-\w_{\gamma})\,,
\end{eqnarray}
where the TE- and TM-polarized fields $\mathbf{E}^{\rm
TE}_{\kpv}(\rhob,z)$ and $\mathbf{E}^{\rm TM}_{\kpv}(\rhob,z)$ are
given by Eq.~(\ref{eq:TE-rad-profs-E0}) and
(\ref{eq:TM-rad-profsE0}), respectively. It is convenient to re-write the emission
rate as a function of the LDoS for radiative states $J_{\rm rad}(\w_{0},z)$ according to
\begin{equation}\label{eq:Gamma_rad2}
    \Gamma(z)=\frac{4\pi^{2}\md^2\w_{0}}{\hbar}J_{\rm rad}(\w_{0},z)\,,
\end{equation}
whith
\begin{eqnarray}\label{eq:density-rad1}
    J_{\rm rad}(\w_{0},z)&=&\frac{S}{(2\pi)^2}\sum_{p=\rm TE,\,TM}\sum_{j=0,\,\rm M+1}\int
    |\mathbf{E}^{p}_{\kpv}(\rhob,z)\cdot\epsd|^{2}\nonumber\\
    &\times&\rho_{j}(\kpv,\w)\ud\kpv\,,
\end{eqnarray}
and $\rho_{j}(\kpv,\w)$ being the one-dimensional (1D) photon DoS
at a fixed in-plane wavevector $\kpv$, for radiative modes outgoing in the medium $j$:
\begin{eqnarray}\label{eq:1D-DOS}
    \rho_{j}(\kpv,\w)&=&\frac{2\w_{0}}{c^2}\sum_{q}\delta\left(\frac{\w_{0}^{2}}{c^2} -
    \frac{\w_{\gamma}^{2}}{c^2}\right)\nonumber\\
    &=&
    \displaystyle\frac{L\sqrt{\eps_j}\w_{0}}{2\pi c}\frac{\Theta(\w_{0}^2-
    \displaystyle\frac{c^{2}\kp^{2}}{\eps_j})}
    {\displaystyle\sqrt{\w_{0}^{2}-\frac{c^{2}\kp^{2}}{\eps_j}}}\,.,
\end{eqnarray}
where $L=V/S$ is the width of the normalization box in the $z$-direction (which disappears in the final expression of the SE rate) and 
$\Theta$ [$\Theta(x)=1\,(=0)$ if $x>0\,(x<0)$] is the Heaviside function.
It is worth to stress that, by using Eq.~(\ref{eq:density-rad1}) with the basis of radiative states
discussed in Sec.~\ref{subsec:rad_modes}, we get rid of any ambiguity in the definition
of the LDoS: for each outgoing radiative mode ($j=0$ or $j=\rm{M+1}$) the LDoS is defined by
a single mode-component only and thus any difficulty related
to interference effects between components of different modes is avoided. Also, due to the Heaviside function
in Eq.~(\ref{eq:1D-DOS}), emission into partially radiative modes occurs only in the cladding
with the higher refractive index.

From Eqs.~(\ref{eq:Gamma_rad2})-(\ref{eq:1D-DOS}), and after the introduction of spherical coordinates in the $(\kpv,q)$ space,
\begin{equation}\label{eq:spherical-coords}
    \kpv=(k_{j}\textrm{sin}\theta\textrm{cos}\phi,\,k_{j}\textrm{sin}\theta\textrm{sin}\phi)\,,
    \quad \phi\,\epsilon [0,\,2\pi]\,,\,\,\theta\,\epsilon [0,\pi/2]\,,
\end{equation}
the single contributions to the total emission rate $\Gamma(z)$ due to the decay of horizontal
and vertical dipoles are easily obtained:
\begin{equation}\label{eq:rate-rad-TE-par}
  \Gamma_{\parallel}^{\rm{TE}}(z)=\frac{\md^2\w_{0}^{3}}
  {2\hbar c^3}\sum_{j=0,\,\rm M+1}\eps_{j}^{3/2}\int_{0}^{\pi/2}
  |E^{\rm TE}(\kp=k_{j}\textrm{sin}\theta,z)|^{2}\rm{sin}\theta\ud\theta\,,\\
\end{equation}
\begin{equation}
\label{eq:rate-rad-TM-par}
  \Gamma_{\parallel}^{\rm{TM}}(z)=\frac{\md^2\w_{0}}
  {2\hbar c[\eps(z)]^{2}}\sum_{j=0,\,\rm M+1}\eps_{j}^{3/2}\int_{0}^{\pi/2}
  |E_{\parallel}^{\rm TM}(\kp=k_{j}\textrm{sin}\theta,z)|^{2}\rm{sin}\theta\ud\theta\,,\\
\end{equation}
\begin{equation}
\label{eq:rate-rad-TM-perp}
  \Gamma_{\perp}^{\rm{TM}}(z)=\frac{\md^2\w_{0}}
  {\hbar c[\eps(z)]^{2}}\sum_{j=0,\,\rm M+1}\eps_{j}^{3/2}\int_{0}^{\pi/2}
  |E_{\perp}^{\rm TM}(\kp=k_{j}\textrm{sin}\theta,z)|^{2}\rm{sin}\theta\ud\theta\,,\\
\end{equation}
where the field amplitudes $E^{\rm TE}$, $E^{\rm TM}_{\parallel}$ and $E^{\rm TM}_{\perp}$ are given
by Eqs. (\ref{eq:TE-rad-profs-E1}), (\ref{eq:TM-rad-profs-E2}) and (\ref{eq:TM-rad-profs-E1}), respectively.
\subsection{Emission rates into guided modes}
According to Eq.~(\ref{eq:golden_rule2}), the spontaneous emission rate for the
decay into guided modes having frequency $\wg$ is given by
\begin{eqnarray}\label{eq:Gamma_gui}
    \Gamma(z)&=&\frac{4\pi^{2}\md^{2}}{\hbar}\sum_{p=\textrm{TE,TM}}\sum_{\alpha}
    \sum_{\kpv}\mid\mathbf{E}_{\kpv}^{p}
    (\rhob,z)\cdot\epsd\mid^{2}\wg\delta(\omega_{0}-\wg)\\\nonumber
    &=&\frac{4\pi^{2}\md^2\w_{0}}{\hbar}J_{\rm gui}(\w_{0},z)\,,
\end{eqnarray}
where the fields $\mathbf{E}_{\kpv}^{\rm TE}(\rhob,z)$
and $\mathbf{E}_{\kpv}^{\rm TM}(\rhob,z)$ are given by
Eq.~(\ref{eq:TE-guided-profs-E0}) and Eq.~(\ref{eq:TM-guided-profs-E0}),
respectively, the sum extends over all the $\alpha$ guided modes, and
the 2D LDoS $J_{\rm gui}(\w_0,z)$ is given by
\begin{equation}\label{density-gui1}
    J_{\rm gui}(\w_0,z)=\frac{S}{(2\pi)^2}\sum_{p=\rm TE,\,TM}
    \sum_{\alpha}\int|\mathbf{E}_{\kpv\alpha}^{p}(\rhob,z)
    \cdot\epsd|^{2}\ud\kpv\,.
\end{equation}
The emission rates $\Gamma_{\parallel}^{\rm TE}$, $\Gamma_{\parallel}^{\rm TM}$ and
$\Gamma_{\perp}^{\rm TM}$ can be easily obtained after integration over $\kpv$
of Eq.~(\ref{eq:Gamma_gui}):
\begin{equation}\label{eq:rate-gui-TE-par}
    \Gamma_{\parallel}^{\rm TE}(z)=\frac{\md^{2}\pi\wo^{3}}{\hbar c^2}\sum_{\alpha}\mid
    E^{\rm TE}(\kp=k_{0}^{\alpha},z)\mid^2\frac{k_{0}^{\alpha}}{v_{0}^{\alpha}}\,,
\end{equation}
\begin{equation}\label{eq:rate-gui-TM-par}
    \Gamma_{\parallel}^{\rm TM}(z)=\frac{\md^{2}\pi c^{2}}{\hbar \w_{0}}\sum_{\alpha}\mid
    E_{\parallel}^{\rm TM}(\kp=k_{0}^{\alpha},z)\mid^2\frac{k_{0}^{\alpha}}{v_{0}^{\alpha}}\,,
\end{equation}
\begin{equation}\label{eq:rate-gui-TM-perp}
    \Gamma_{\perp}^{\rm TM}(z)=\frac{\md^{2}2\pi c^{2}}{\hbar \w_{0}}\sum_{\alpha}\mid
    E_{\perp}^{\rm TM}(\kp=k_{0}^{\alpha},z)\mid^2\frac{k_{0}^{\alpha}}{v_{0}^{\alpha}}\,,
\end{equation}
where $E^{\rm TE}$, $E_{\parallel}^{\rm TM}$ and $E_{\perp}^{\rm TM}$ are given by
Eqs.~(\ref{eq:TE-guided-profs-E1}), (\ref{eq:TM-guided-profs-E2}) and (\ref{eq:TM-guided-profs-E1}),
respectively. In the expressions given above, $k_{0}^{\alpha}=\kp^{\alpha}(\w=\wo)$ and
$v_{0}^{\alpha}=(d\wg/d\kp)_{\wg=\w_{0}}$ are the in-plane
wavevector and the group velocity of the $\alpha-$th guided
mode calculated at the dipole emission frequency $\omega_{0}$,
respectively. The wavevectors $k_{0}^{\alpha}$ as functions of the frequencies
can be found as the poles (which are real ones for guided modes) of the transmission amplitude
$t=1/\mathrm{T}_{22}$ of the whole dielectric
structure, $\mathrm{T}$ being the total transfer matrix.

\section{Applications}
In this Section we apply the formalism previously developed in order
to investigate the SE process in realistic multilayer
structures. As a typical high-index dielectric material, we take Silicon (n$_{\rm Si}$=3.48).
After the analysis of a single Silicon/air interface, we
will examine and compare the emission and confinement properties of
different Silicon waveguides, namely a standard waveguide slab
consisting of a Silicon core surrounded by two cladding layers with
the same refractive index (symmetric configuration) or different
ones (asymmetric configuration) and the silicon-on-insulator Slot
waveguide. The SE rate has been evaluated for dipoles emitting at
$\lambda_0=\w_0/c=1.55\,\mu\textrm{m}$ which is the typical emission
wavelength of Erbium ions (Er$^{3+}$) often used as the active layer
of Silicon-based light sources (see e.g. the review paper by
Kenyon~\cite{Kenyon05}). All the rates shown have been normalized
with respect to the vacuum emission rate
$\Gamma=\Gamma_0=(4\md^{2}\w_{0}^3)/(3\hbar c^3)$ of a randomly
oriented dipole.
\begin{center}
\begin{figure*}[t!]
\resizebox{\hsize}{!}{
  \includegraphics*{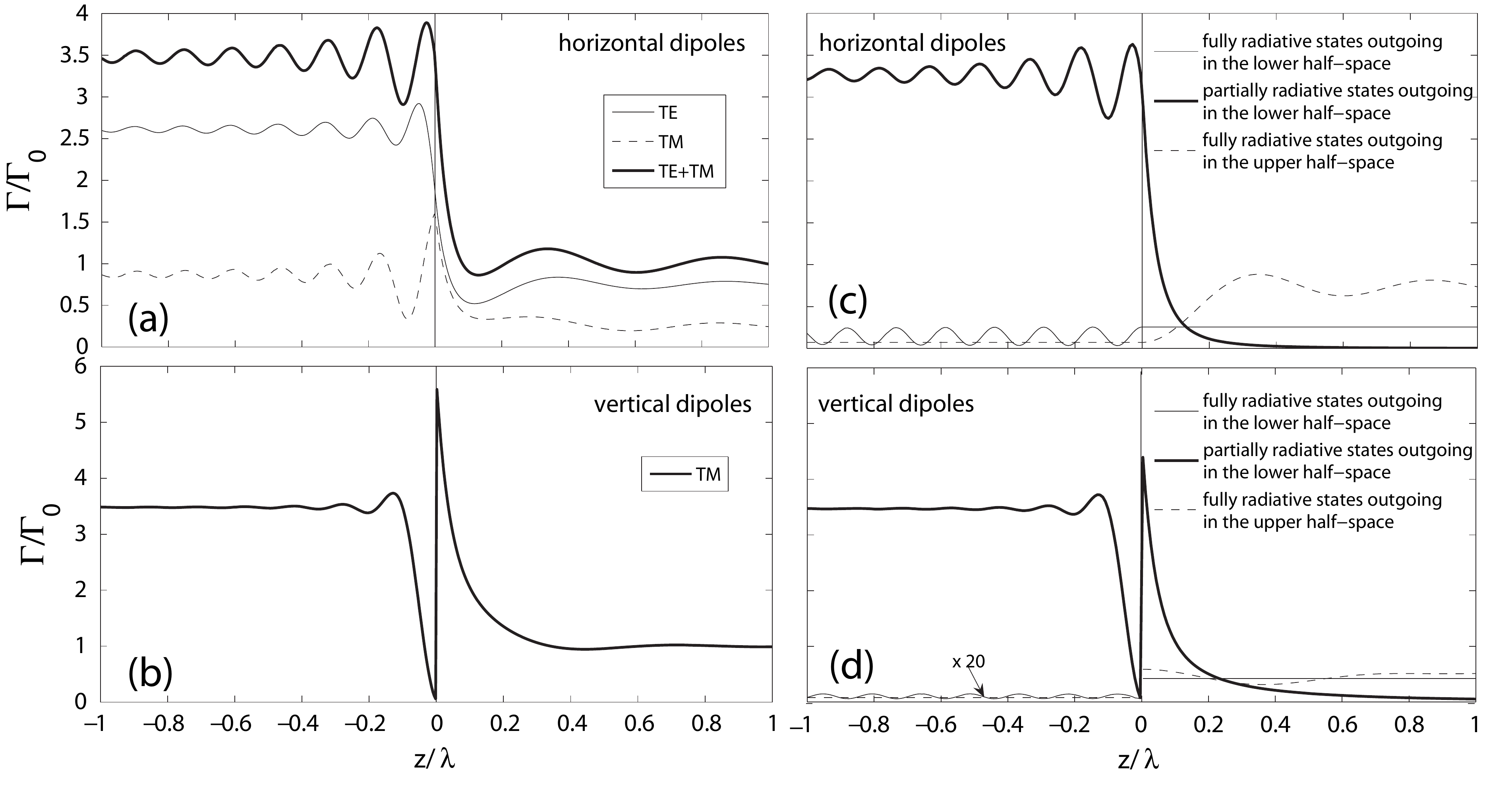}}\\
  \caption{The normalized spontaneous emission rate as function of the dipole position $z/\lambda$ ($\lambda=\lambda_0$)
  in a dielectric/air interface. The half space $z/\lambda<0$ is made by Silicon with refractive index $n_{\rm Si}=3.48$.
  (a): the contribution from horizontal (in-plane oriented) dipoles decaying  into TE- and TM-polarized modes.
  (b): the contribution from vertical ($\hat{z}$ oriented) dipoles which couple to TM-polarized modes only. (c) and (d):
  the total emission rates due to fully and partially radiative modes for horizontal and vertical dipoles.} \label{fig:interface}
\end{figure*}
\end{center}
\begin{center}
\begin{figure*}[t!]
\resizebox{\hsize}{!}{
  \includegraphics*{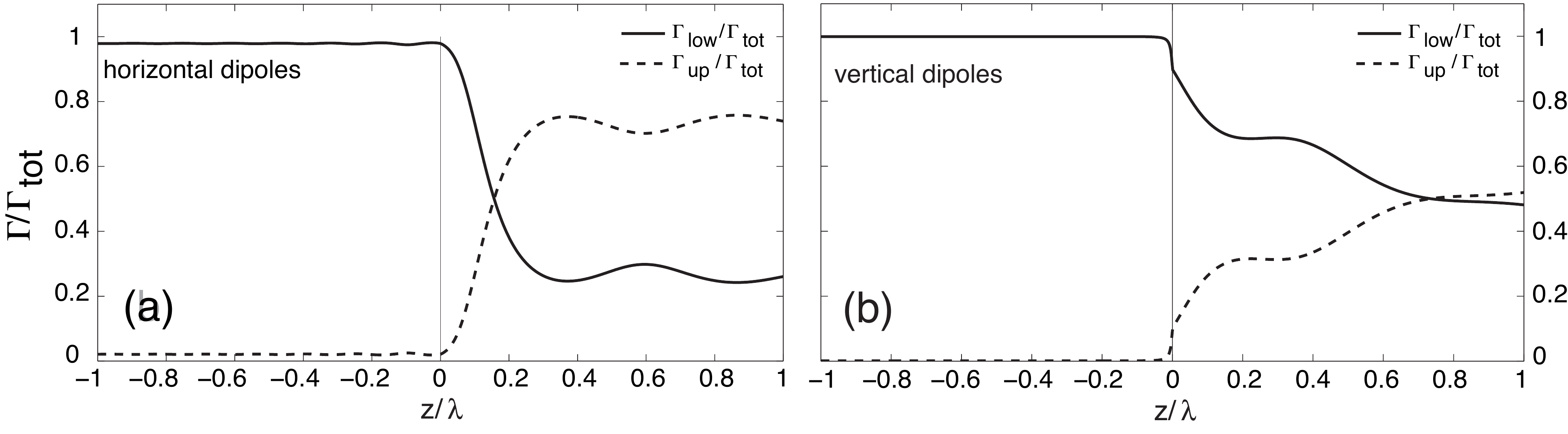}}\\
  \caption{The spontaneous emission rates $\Gamma=\Gamma_{\rm low}$ and $\Gamma=\Gamma_{\rm up}$, 
  for light outgoing in the lower and upper half-space, respectively, as function of the dipole position $z/\lambda$ for
  the same dielectric/air interface depicted in Fig.~\ref{fig:interface}. The rates are normalized to the total emission rate 
  $\Gamma_{\rm tot}=\Gamma_{\rm low}+\Gamma_{\rm up}$.} \label{fig:interface-contributions}
\end{figure*}
\end{center}
Figure~\ref{fig:interface} shows the normalized spontaneous emission
$\Gamma/\Gamma_0$ for a Silicon/air interface as function of
$z/\lambda$. The emission rate for horizontal dipoles decaying into
TE- and TM-polarized modes [see Fig.~\ref{fig:interface}(a)] varies
continuously through the interface, as required from the continuity
condition of the tangential field component at a dielectric
boundary, while the emission rate for vertical dipoles (which couple
to TM-polarized modes only), is discontinuous at the same point [see
$z=0$ in Fig.~\ref{fig:interface}(b)] due to the discontinuity of
the $z$ component of the electric field. Far from the interface
boundary, when $z/\lambda\gg 1$, the spontaneous emission rate (for
both horizontal and vertical dipoles) is scaled by the refractive
index according to $\Gamma(z)=\sqrt{\eps(z)}\Gamma_0$, in agreement
with earlier works~\cite{Glauber91,Loudon91,Zakowicz94}, with
oscillations around the average value. The contributions to the
total emission rate due to the decay into partially and fully
radiative modes are shown in Figs.~\ref{fig:interface}(c)-(d): in
the dielectric half space the emission is mainly due to the
partially radiative states (see the thick solid lines for $z/\lambda
< 0$) which also characterize the profile of the total emission rate
in the proximity of the interface boundary [see the thick solid
lines at values $z/\lambda$ between 0 and 0.2 in
Figs.~\ref{fig:interface}(a)-(c) and in
Figs.~\ref{fig:interface}(b)-(d)] and decay exponentially (in form
of evanescent waves) in the free half space far from it. Hence, the
evanescent component of partially radiative modes (which does not
contribute to the total energy flux and it is hidden in standard far
field experiments) turns out to be relevant in radiation emission
analysis, since it strongly affects the radiative lifetime
$\tau=1/\Gamma$ in the vicinity of the interface boundary. 
Moreover, for vertical dipoles in a generic dielectric/air interface,
one can analytically work out that, in the limit of  a very large refractive index $n\gg\,1$, 
the emission into partially radiative modes (which is the dominant one) at the discontinuous boundary, 
is given by $\Gamma(z\rightarrow\,0^{-})/\Gamma_0=1/n^3$ and $\Gamma(z\rightarrow\,0^{+})=n$, 
in agreement with an earlier work by Loudon~\cite{Loudon91,note_gamma}. 

The contributions to the total SE rate 
corresponding to light emitted either in the lower or in the upper layer are shown in 
Fig.~\ref{fig:interface-contributions} as a function of the dipole position. It is worth to notice that, within our model, 
these quantities are straightforwardly obtained by selecting 
the single outgoing radiative mode (see Fig.~\ref{fig:rad-modes} in Sec.~\ref{sec:modes}), 
through the index-layer $j=0$ (emission into the lower cladding) or $j=M+1$ (emission into the upper cladding) 
in Eq.~(\ref{eq:density-rad1}). 
These rates could also be obtained by using the standard basis with a single ingoing and two outgoing components~\cite{Carniglia71}, 
but in this case the interference terms between the two modes of Fig.2a must be explicitly calculated~\cite{Zakowicz95a}.
 Thus, the present approach using the basis with a single outgoing component 
is especially useful for calculating radiative patterns and the emitted light in the lower/upper half spaces, 
which is a physically and technologically important problem for light emitting structures like LEDs 
and vertical laser diodes.

\begin{center}
\begin{figure}[htb!]
  \includegraphics[height=11cm]{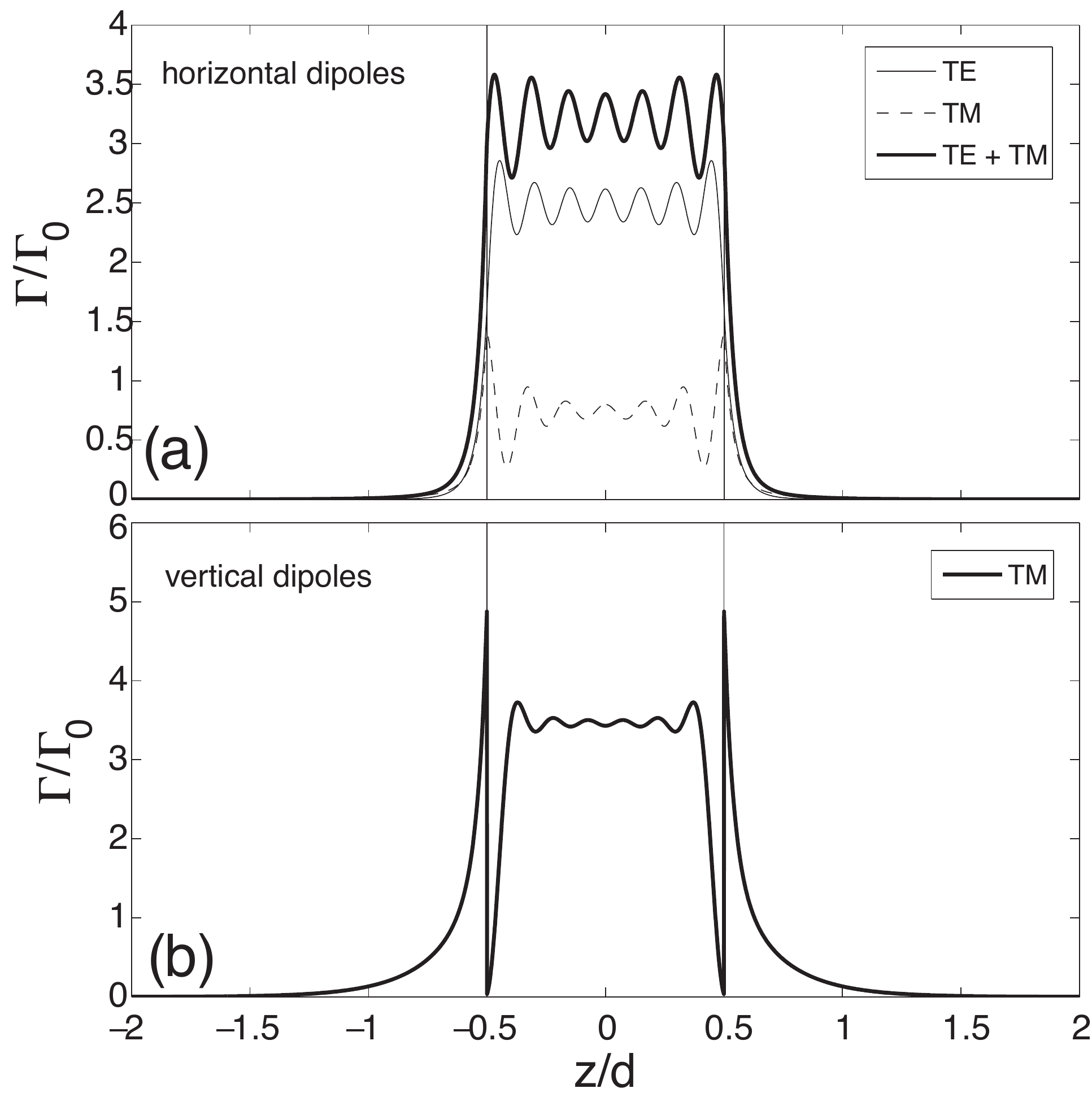}\\
  \caption{The normalized spontaneous emission
  rate into guided modes for a symmetric Silicon waveguide with air claddings as a function of $z/d$,
  $d$ being the thickness of the Silicon core. The refractive index in the half spaces
  $|z|>d/2$ is 1 and the core thickness has been taken equal to vacuum emission
  wavelength $\lambda_0$. (a): the contribution of horizontal dipoles.
  (b): the contribution of vertical dipoles.}\label{fig:Sigui1l}
\end{figure}
\end{center}

\begin{center}
\begin{figure}[htb!]
  \includegraphics[height=11cm]{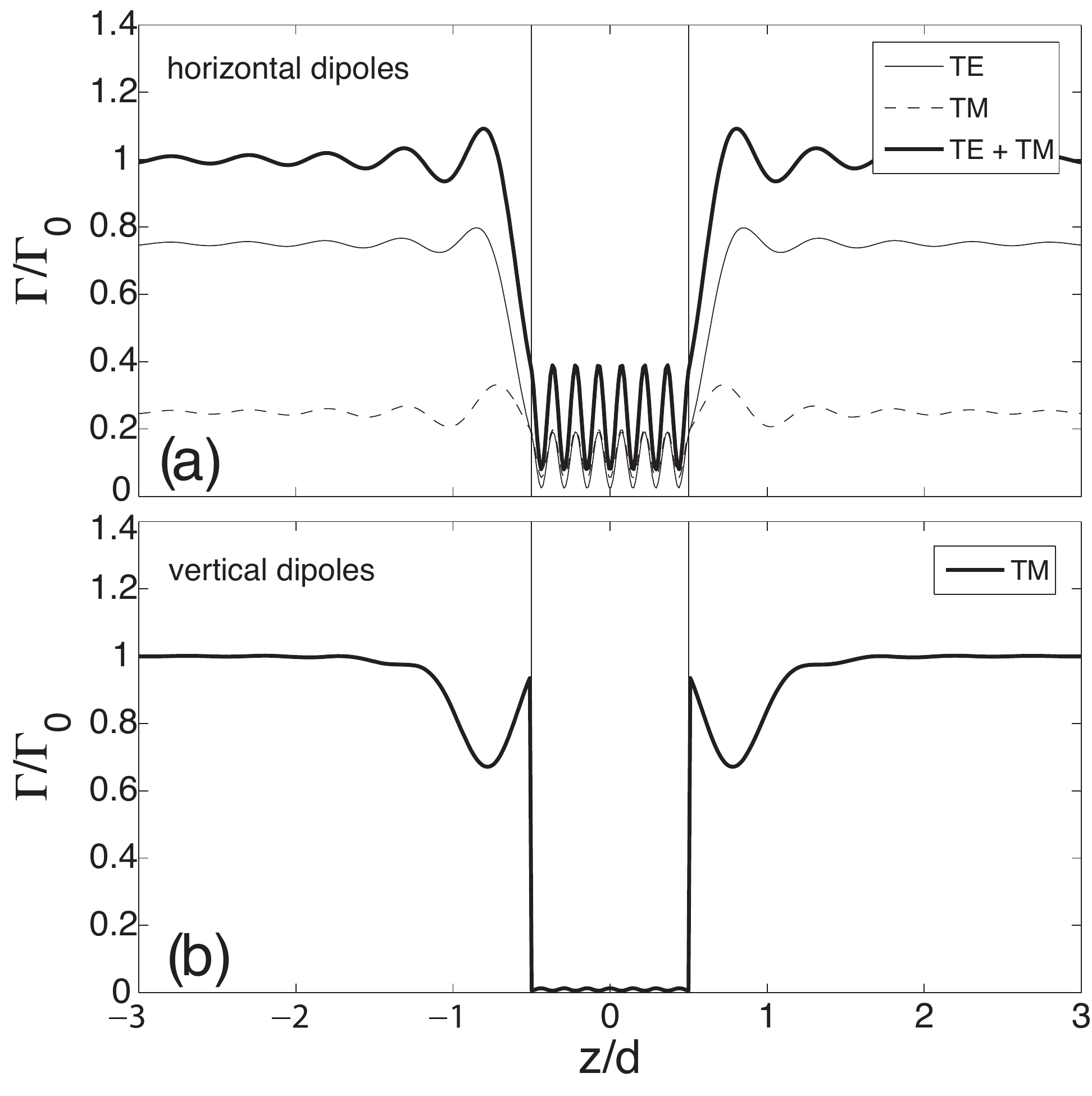}\\
  \caption{The normalized spontaneous emission
  rate into radiative modes for a symmetric Silicon waveguide with air claddings as a function of $z/d$,
  The refractive index in the half spaces $|z|>d/2$ is 1 and the core thickness
  has been taken equal to vacuum emission
  wavelength $\lambda_0$. (a): the contribution of horizontal dipoles.
  (b): the contribution of vertical dipoles.}\label{fig:Sirad1l}
\end{figure}
\end{center}

\begin{center}
\begin{figure}[htb!]
  \includegraphics[height=12cm]{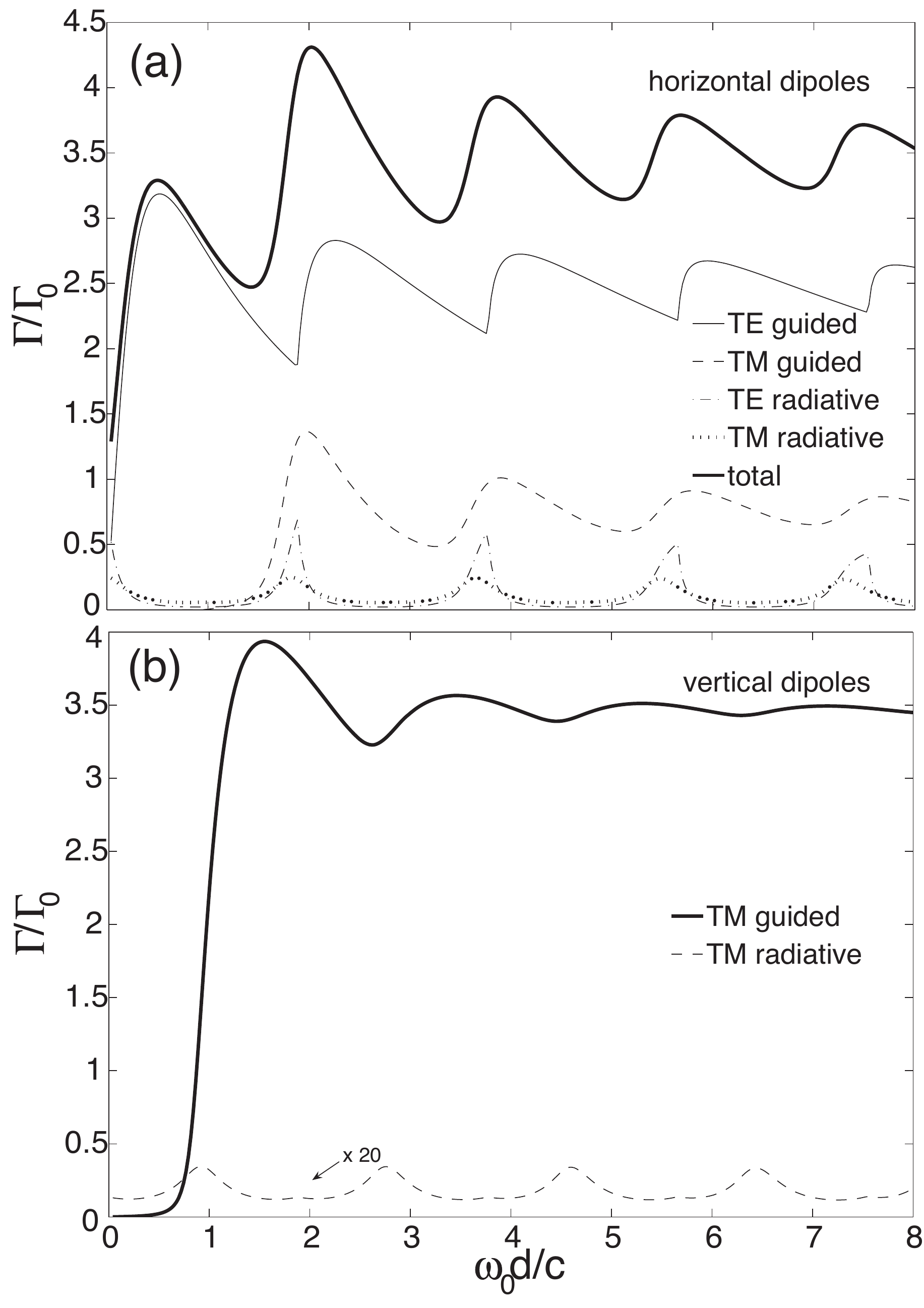}\\
  \caption{The normalized spontaneous emission rate for a symmetric Silicon
  waveguide with air claddings evaluated as a function of the photonic thickness $\displaystyle (\wo\,d)/c$ and
  for a dipole kept at the middle of the Silicon core.
  (a): the emission from horizontal dipoles.
  (b): the emission from vertical dipoles.}\label{fig:si-comparison}
\end{figure}
\end{center}

The spontaneous emission rates for a symmetric Silicon waveguide are
shown in Figs.~\ref{fig:Sigui1l} and \ref{fig:Sirad1l} as functions
of $z/d$, where $d$ is the thickness of the Silicon core ($|z|< d/2$)
surrounded by air. Such a structure supports a finite number of
guided modes and, since the upper ($z>d/2$) and lower ($z<d/2$)
claddings have the same refractive index, only fully radiative modes
can be excited and propagate out from the waveguide. By choosing a
thickness $d=\lambda_0$, one can calculate 7 TE and TM guided modes
whose contribution to the total emission rate is significantly
greater than the contribution due to the emission into radiative
modes, as it can be seen by comparison of
Figs.~\ref{fig:Sigui1l}(a)-(b) with Figs.~\ref{fig:Sirad1l}(a)-(b).
Furthermore, the emission rate in the core [see the spatial range
$-1/2<z/d<1/2$ in Figs.~\ref{fig:Sigui1l}(a)-(b)] is close to the
bulk value $n_{\rm Si}\Gamma_0$.
The influence of an increasing number of guided modes is
investigated in Fig.~\ref{fig:si-comparison}, where the emission
rate has been evaluated as a function of the so called dimensionless
photonic thickness $\w_{0}d/c$, while keeping the dipole position at
the centre ($z=0$) of the waveguide. It can be noticed that, with
increasing thickness $d$, the contribution from the new arising
modes is associated with the appearance of distinct features in the
emission pattern such as dips and peaks.
Furthermore, for vertical dipoles, the spontaneous emission rate is
drastically suppressed for waveguide thickness below $\displaystyle
d=0.5(c/\w_{0})=0.5(\lambda_0/2\pi)$ [see
Fig.~\ref{fig:si-comparison}(a)], whereas, for the same range of
thicknesses, the total emission rate from horizontal dipoles [see
the thick solid line in Fig.~\ref{fig:si-comparison}(b)] is
$\Gamma\approx\,1.3\div 3.3\,\Gamma_0$ and mainly due the excitation
of TE guided modes. Also, for thicknesses $d\gtrsim
2c/\w_{0}=\lambda_{0}/\pi$, the contributions to the total emission
rate due to horizontal and vertical dipoles become comparable and
close to the bulk value $n_{\rm Si}\Gamma_0$.

The above results, which follow from the mode decomposition based on
a single outgoing component for radiative states (see previous
Sec.~\ref{subsec:rad_modes}), are in agreement with those shown
in earlier works~\cite{Loudon92,Urbach98,Ho99} and which have been
obtained by using the standard set of radiative modes based on the
triplets incident-reflected- transmitted waves.
\begin{center}
\begin{figure}[htb!]
  \includegraphics[height=11cm]{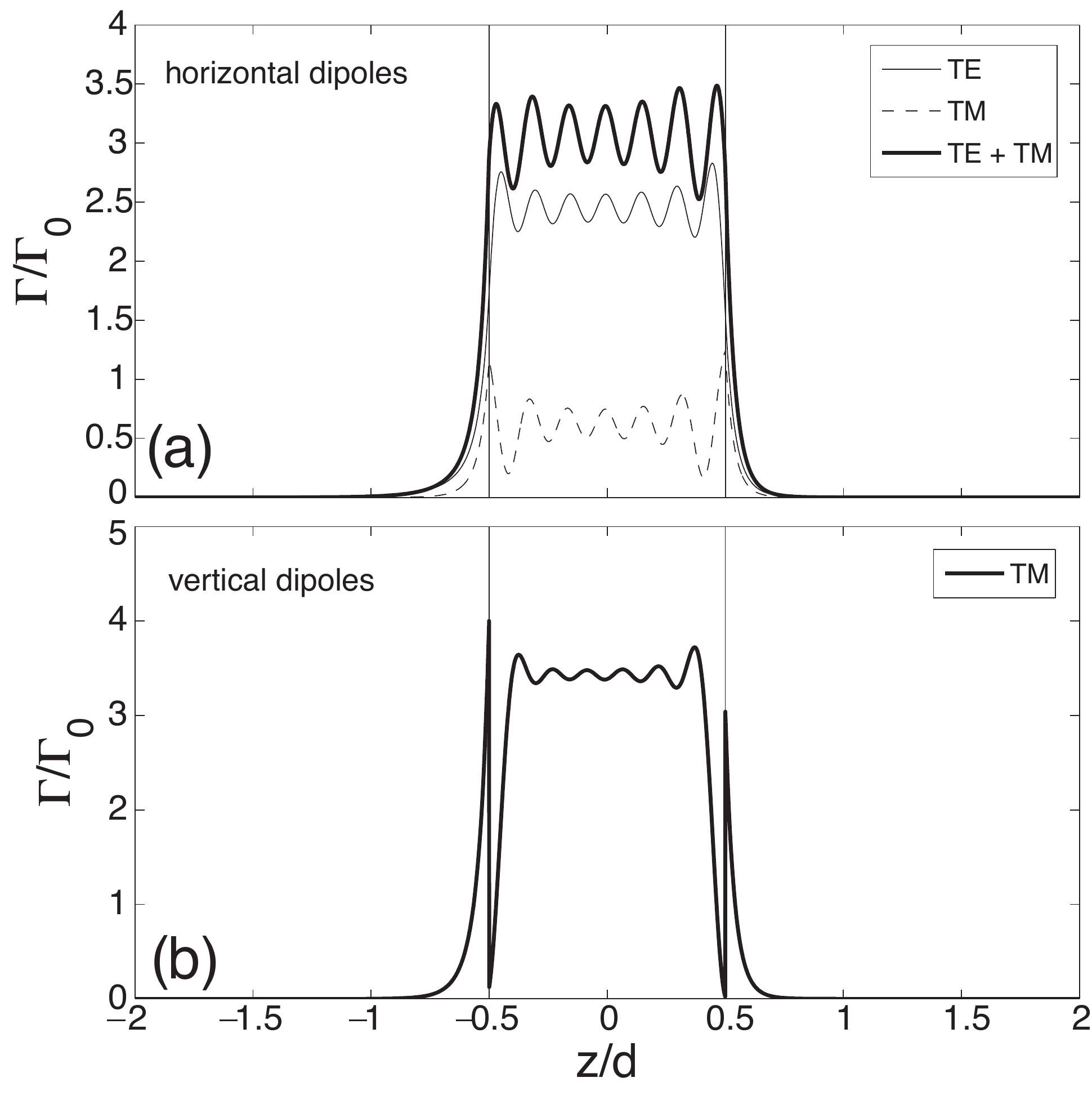}\\
  \caption{The spatial dependence of the normalized spontaneous emission
  rate into guided modes for the asymmetric Silicon waveguide (SiO$_2$/Si/air).
  The thickness $d$ of the Silicon core ($n_{\rm Si}=3.48$) is equal to the vacuum emission wavelength
  $\lambda_0$. The lower cladding ($z<d/2$) is made by SiO$_2$ ($n_{\rm SiO_2}=1.45$) and the
  refractive index of the upper cladding ($z>d/2$) is 1.
  (a): the contribution of horizontal dipoles.
  (b): the contribution of vertical dipoles.}\label{fig:aSigui1l}
\end{figure}
\end{center}
\begin{center}
\begin{figure}[htb!]
  \includegraphics[height=11cm]{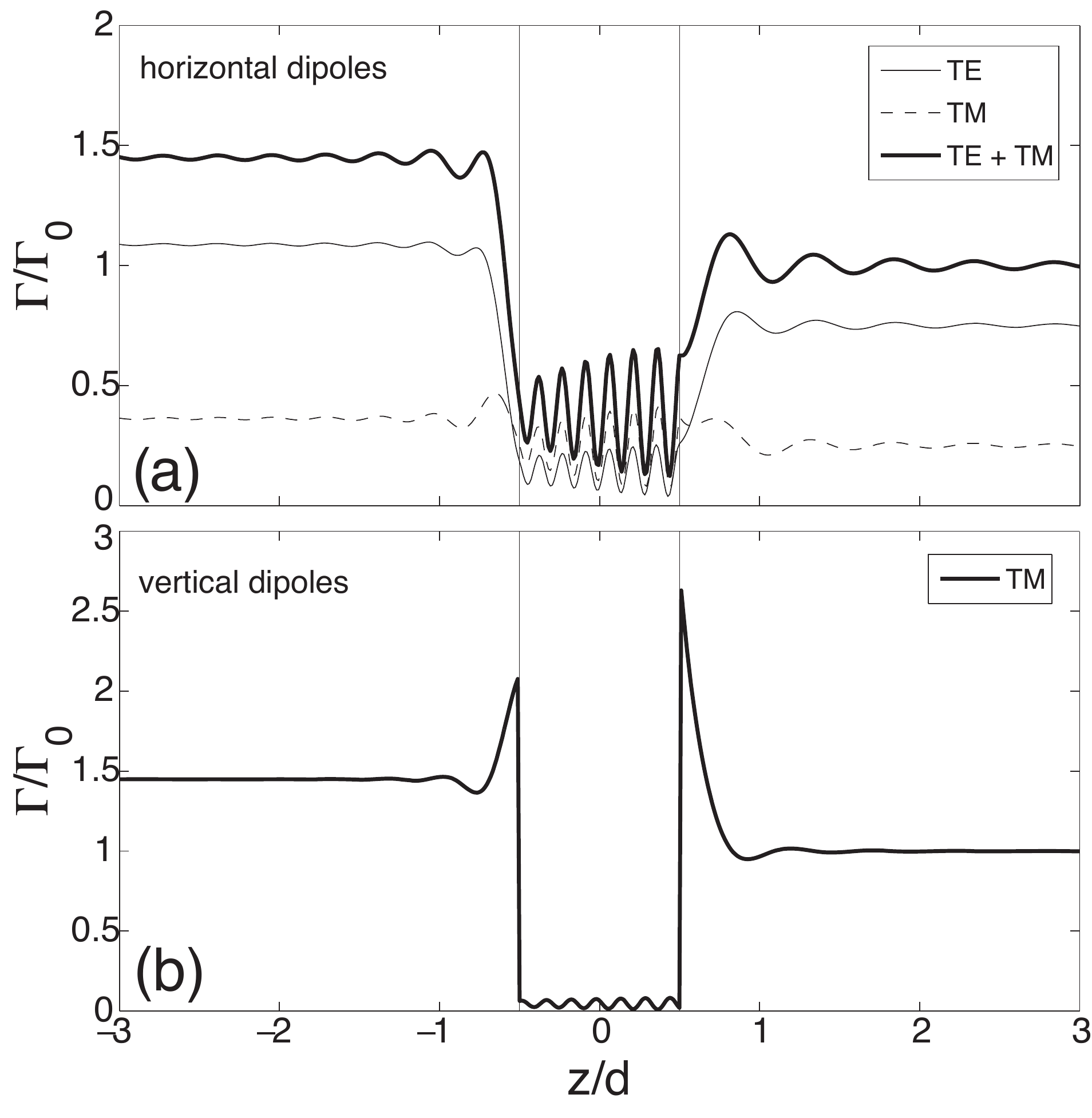}\\
  \caption{The spatial dependence of the normalized spontaneous emission
  rate into radiative modes for the asymmetric Silicon waveguide (SiO$_2$/Si/air).
  The same parameters of Fig.~\ref{fig:aSigui1l} have been used.
  (a): the contribution of horizontal dipoles.
  (b): the contribution of vertical dipoles.}\label{fig:aSirad1l}
\end{figure}
\end{center}
We now apply our model to study the SE in an asymmetric dielectric
waveguide, i.e., a waveguide with different refractive indices in
the lower and upper claddings (both values being of course smaller than
the core one). Due to the asymmetry, the condition for total
internal reflection can be met for incidence angles beyond the
limit one, and also partially radiative modes, which are evanescent
in the lower index cladding, can thus be excited in such a
structure. Figures~\ref{fig:aSigui1l} and ~\ref{fig:aSirad1l} show
the  $z$ dependence of the emission rates into guided and
radiative modes, respectively, for an asymmetric waveguide made by a
Silicon core bounded by a Silicon Oxide (SiO$_2$) lower cladding,
and by air in the upper half space acting as upper cladding. For a thickness
$d=\lambda_0$, there are now 7 TE and 6 TM guided modes and the
asymmetry-induced modifications in the emission pattern are clearly
seen, especially for emission into radiative modes. Figure~\ref{fig:aSi-comparison} shows the emission rate as a
function of the core thickness: there are no guided modes for
thicknesses smaller than $d\approx 0.42\,c/\w_0$ and the emission
rate is thus sustained by radiative modes only (see the continuous
thick line). For $d\geq c/\w_0=\lambda_0/2\pi$ the emission is
mainly due to guided modes, the contribution from TE polarized modes being larger.
However, as a consequence of slab asymmetry leading to partially radiative modes, the
contribution from radiative states is larger than in the symmetric waveguide case.
\begin{center}
\begin{figure}[h!]
  \includegraphics[height=12cm]{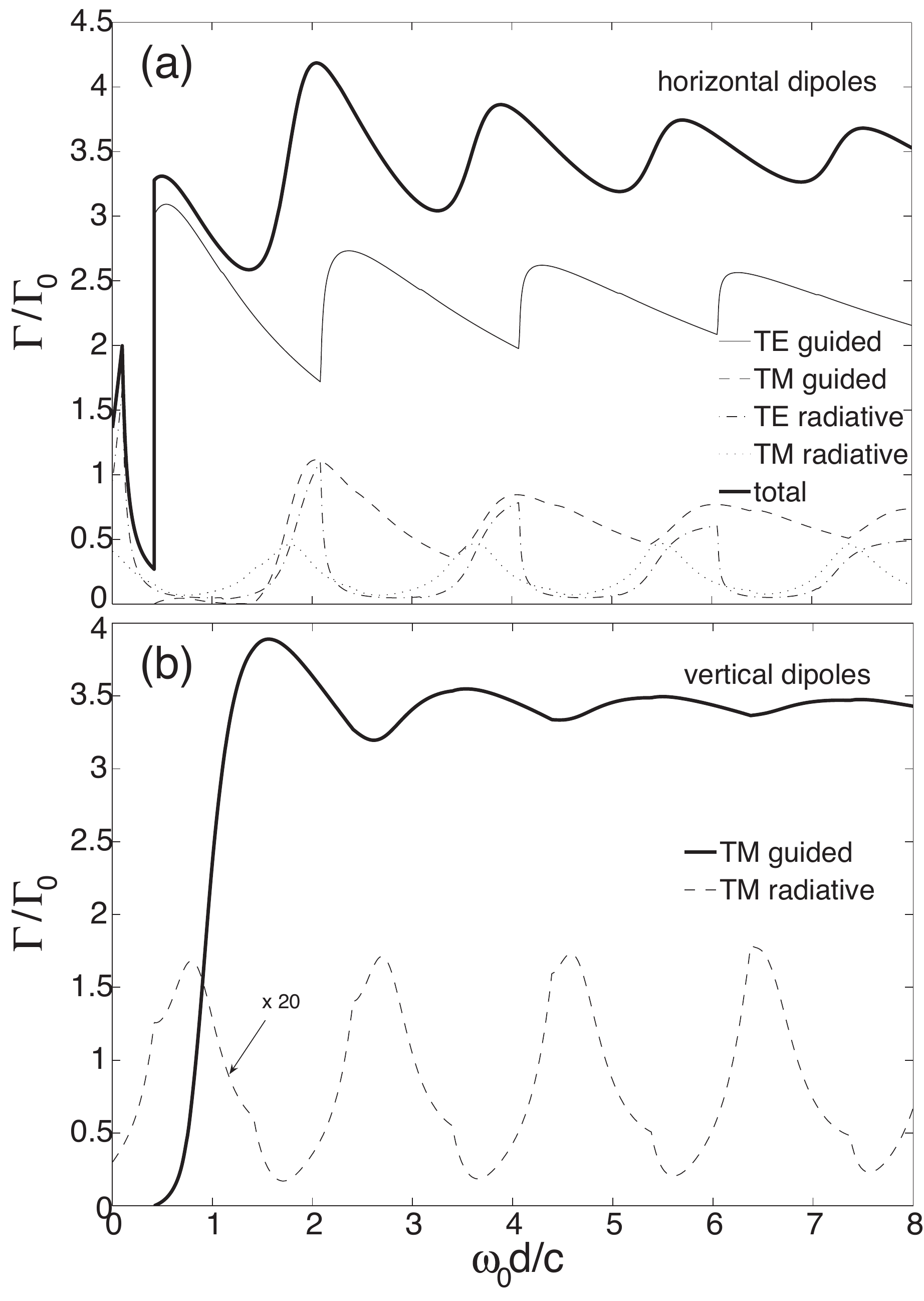}\\
  \caption{The normalized spontaneous emission rate for the asymmetric Silicon
  waveguide (SiO$_2$/Si/air) evaluated as a function of the photonic thickness $\displaystyle (\wo\,d)/c$ and
  for a dipole kept at the middle of the Silicon core.
  The same parameters of Figs.~\ref{fig:aSigui1l} and~\ref{fig:aSirad1l}
  have been used. (a): emission from horizontal dipoles.
  (b): emission from vertical dipoles. For both orientations, the onset of
  emission into guided modes occurs at the threshold value $\omega_{0} d/c\approx 0.42$.}
  \label{fig:aSi-comparison}
\end{figure}
\end{center}
\begin{center}
\begin{figure}[t!]
  \includegraphics[height=9cm]{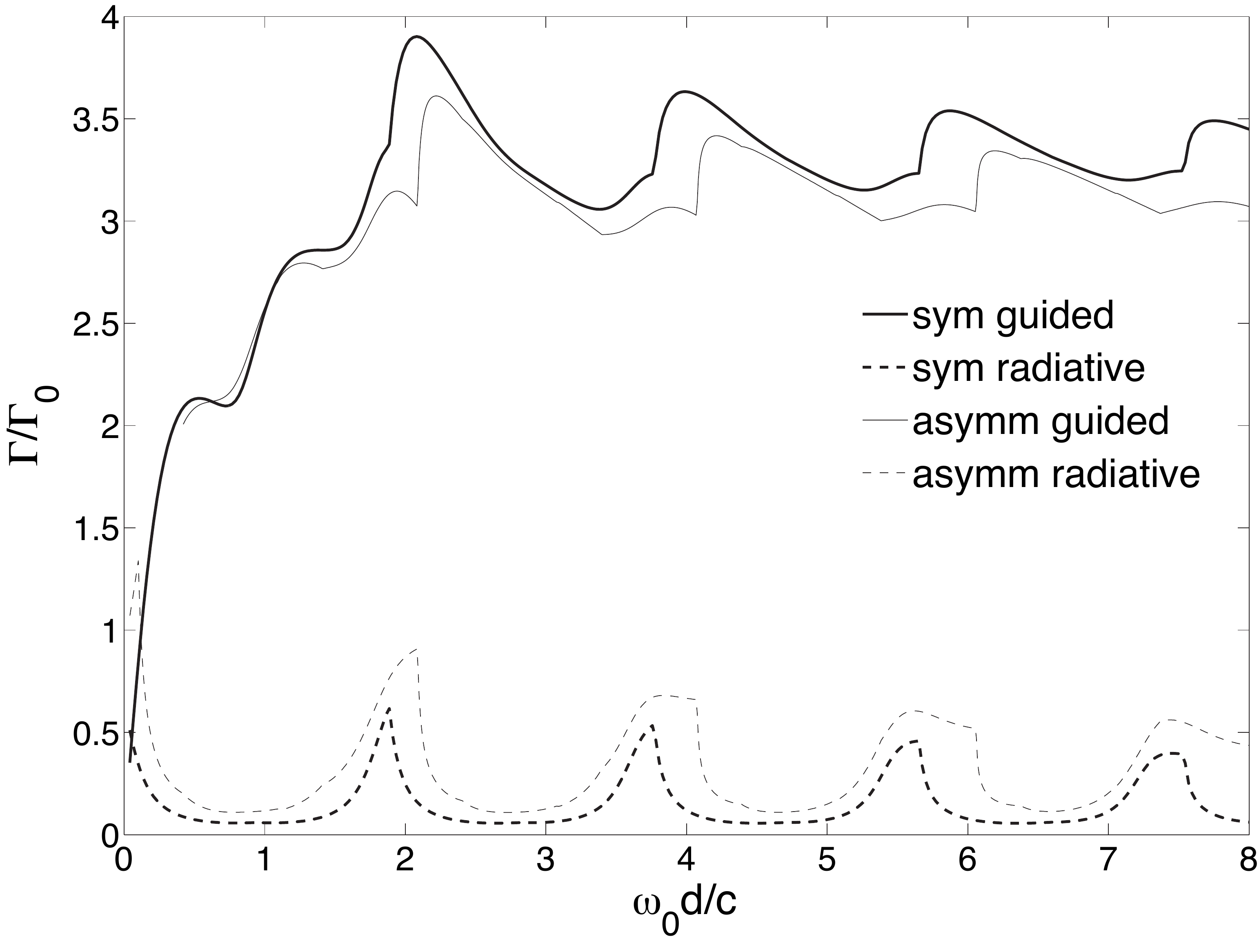}\\
  \caption{The contributions to the total normalized SE rate $\displaystyle\Gamma=\frac{2}{3}\Gamma_{\parallel}
  + \frac{1}{3}\Gamma_{\perp}$ from a randomly oriented dipole into guided and radiative modes,
  for both symmetric (air/Si/air) and asymmetric (SiO$_2$/Si/air) Silicon waveguides as a function of the dimensionless
  thickness $\omega_{0} d/c$.}\label{fig:si-asi}
\end{figure}
\end{center}

Figure~\ref{fig:si-asi} shows the averaged SE rate
$\Gamma=(2/3)\Gamma_{\parallel}+(1/3)\Gamma_{\perp}$ into guided and
radiative modes for both the symmetric and
asymmetric configurations previously analyzed, as a function of the
core thickness. Here $\Gamma_{\parallel}=\Gamma^{\rm TE}_{\parallel}+
\Gamma^{\rm TM}_{\parallel}$ is the sum over the two polarizations for a
planar (horizontal) dipole, while $\Gamma_{\perp}=\Gamma^{\rm TM}_{\perp}$ for
a vertical dipole and the contributions from horizontal ($\Gamma_{\parallel}$)
and vertical ($\Gamma_{\perp}$) dipoles have been averaged as in the
realistic case of a randomly oriented dipole in Si. Again, it can be seen that in
the asymmetric slab case, the contribution of radiative over guided modes in the SE
is increased, mainly due to the fact that the asymmetric slab supports partially
radiative modes that contribute to SE and are taken into account explicitly
in the present calculation.

\begin{center}
\begin{figure}[ht!]
  \includegraphics[height=7cm]{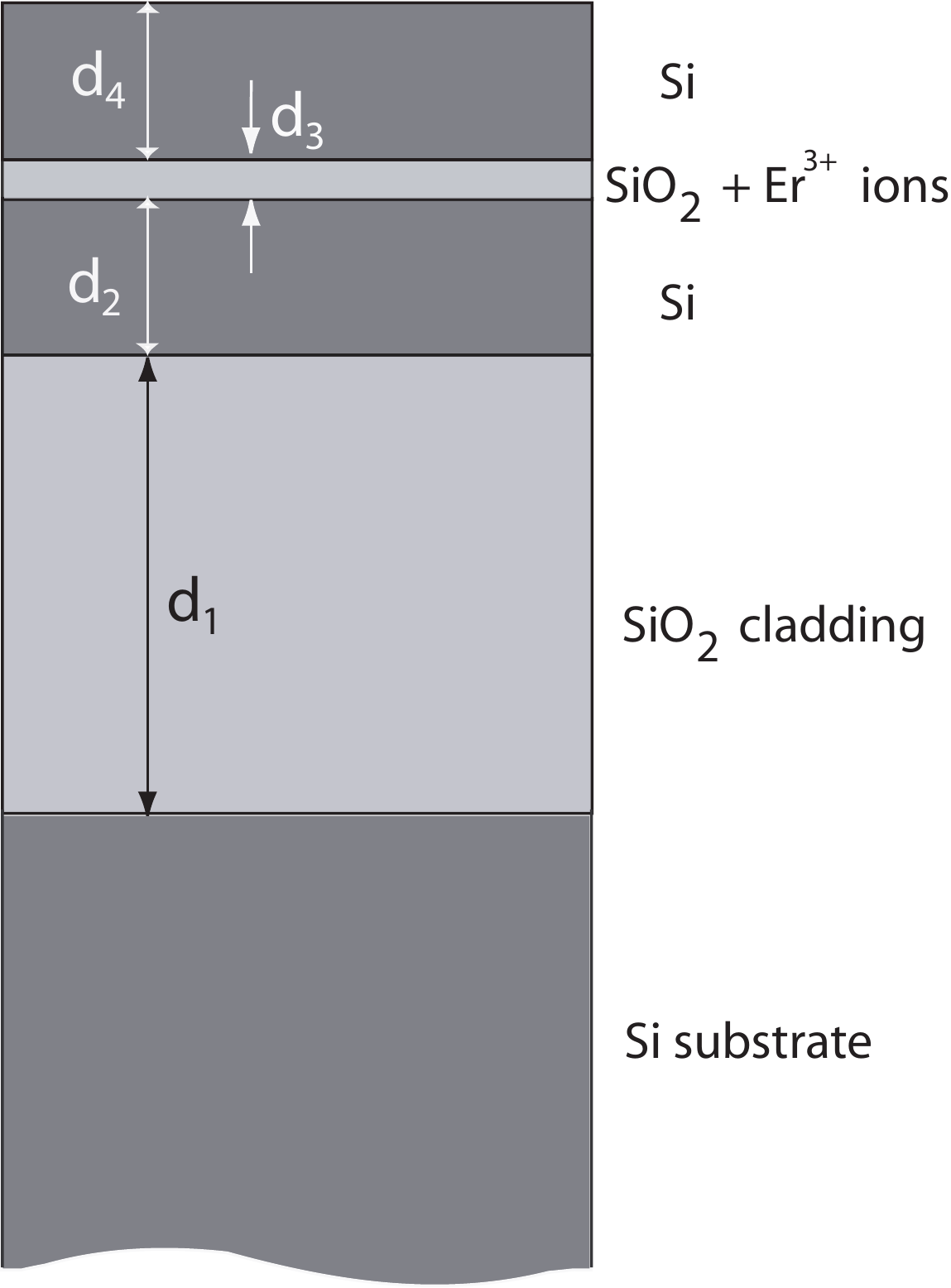}\\
  \caption{Schematic of a Slot waveguide. The core consists of a thin slot of
  Er$^{3+}$-doped SiO$_2$ having thickness d$_3$, embedded between two
  d$_2$- and d$_4$-thick Si layers; the d$_1$-thick SiO$_2$ lower cladding is grown
  on a Si substrate. On top of the last Si layer there is air and thus the numerical
  evaluation of the SE rates has been performed for a 6-layers model structure.
  The following values have been used for the layers thicknesses:
  d$_1$=1.9 $\mu$m, d$_2$=d$_4$=100 nm and d$_3$=20 nm. The values of the refractive
  indices are those which have been used in the structures previously studied.}
  \label{fig:slot-schematic}
\end{figure}
\end{center}

\begin{center}
\begin{figure}[htb!]
  \includegraphics[height=13cm]{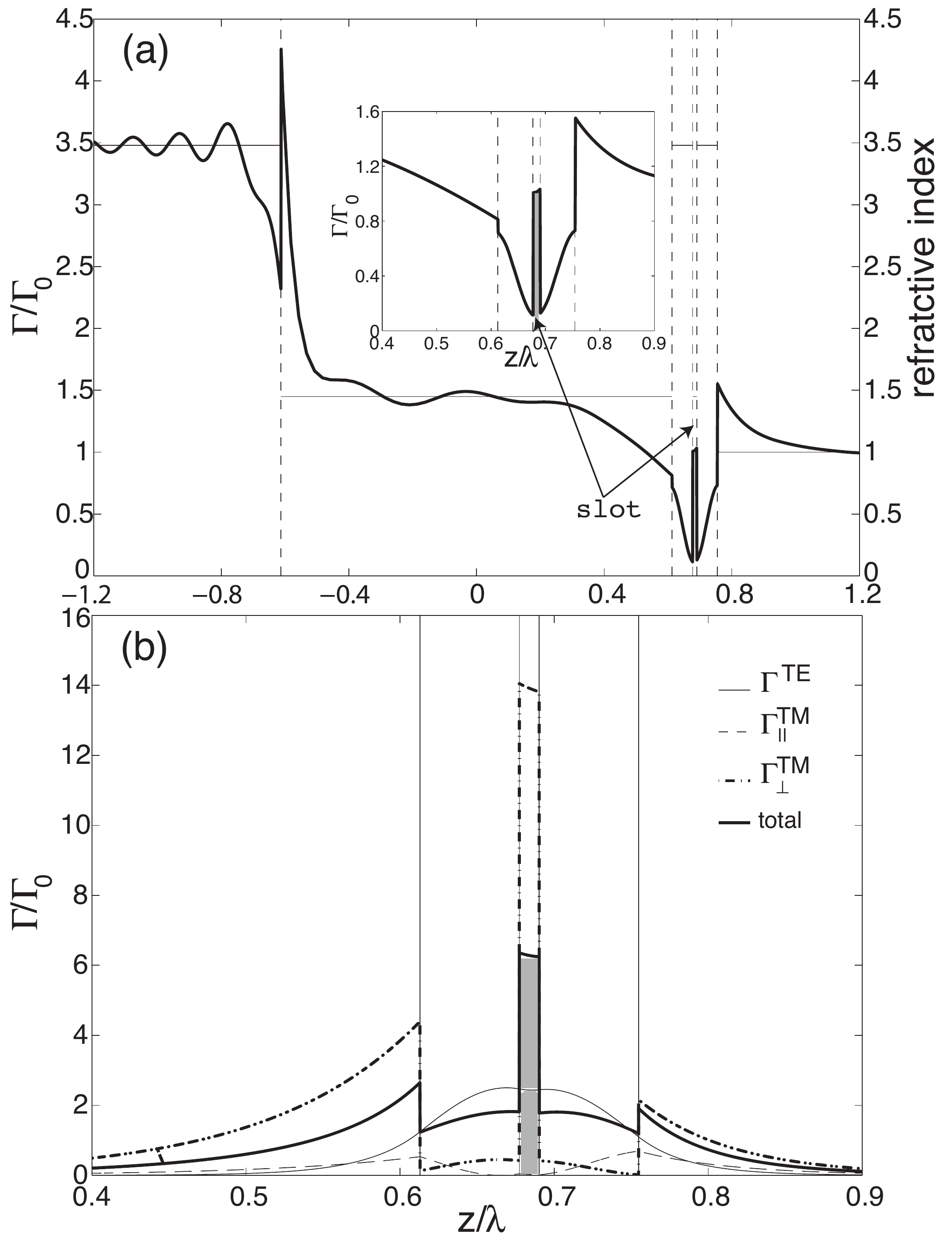}\\
  \caption{The normalized spontaneous emission rate $\Gamma=(2/3)\Gamma_{\parallel}+(1/3)\Gamma_{\perp}$
  for a Slot waveguide
  evaluated as a function of the dipole position.
  (a): the total emission into radiative modes; the refractive index profile is also shown.
  (b): the total emission into guided modes together
  with the separate contributions of both horizontal dipoles decaying into TE
  ($\Gamma^{\rm TE}_{\parallel}=\Gamma^{\rm TE}$)
  and TM modes ($\Gamma_{\parallel}^{\rm TM}$), and vertical dipoles decaying as TM modes only
  ($\Gamma_{\perp}^{\rm TM}$). The layers thicknesses are those reported
  in the caption of Fig.~\ref{fig:slot-schematic}.}
  \label{fig:slot-emission}
\end{figure}
\end{center}

A stronger confinement effect can be achieved in a SOI
slot waveguide. The core of such a waveguide (see the schematic in Fig.~\ref{fig:slot-schematic})
is made up of a very thin layer (slot) of low refractive index active material
(few tens of nanometers thick) embedded between two high-index material regions.
In the configuration here considered, the core consists of a sequence of  Si/SiO$_2$
: Er$^{3+}$/Si layers and lies on the top of a SiO$_2$ cladding grown on a Si
substrate. The discontinuity of
the normal component of the electric field at the high-index-contrast interfaces
of the slot gives rise to an increase of the LDoS, which in turn leads to an enhancement of
SE rate into the waveguide modes.
In Fig.~\ref{fig:slot-emission} the calculated SE rates
$\Gamma=(2/3)\Gamma_{\parallel}+(1/3)\Gamma_{\perp}$ into radiative
and guided modes for a Si Slot waveguide are shown as a function of the dipole position $z/\lambda$
(even if, in a practical case, the Er$^{3+}$ emitters are located in the thin SiO$_2$ layer).
The effect of the discontinuity in the $z$-component
of the electromagnetic field at the slot interfaces can be clearly seen:
the SE rate is mostly due to the decay of vertical ($\hat{z}$-oriented) dipoles into TM guided modes
[see the dashed-dotted line $\Gamma=\Gamma_{\perp}^{\rm TM}$ in Fig.~\ref{fig:slot-emission}(b)],
and the total emission into guided modes is about six times bigger than
the corresponding emission into radiative modes
[see the shaded regions in Figs.~\ref{fig:slot-emission}(a) and \ref{fig:slot-emission}(b)].
Furthermore, after comparison with Fig.~\ref{fig:si-asi}, it is evident
that the light confinement is definitely more effective in a such a Slot waveguide
than in a symmetric Si waveguide of any core-thickness.
Moreover, the calculated SE rates shown in Fig.~\ref{fig:slot-emission}
allow to interpret the experimental results
reported in Ref.~\cite{Galli06} for the enhancement in the photoluminescence from TM
over TE polarized modes for a Slot waveguide containing Er$^{3+}$ ions in the oxide (slot)
layer. The vertical structure is the one depicted in our Fig.~\ref{fig:slot-schematic},
with the same thickness parameters and the emission wavelength is 1.54 $\mu$m.
In the experiment, the TM/TE intensity ratio for light emitted from the edge
of the waveguide is between 6 and 7.5, with a slight dependence on the
position of the excitation spot. From Fig.~\ref{fig:slot-emission}, the calculated TM/TE ratio
for a dipole embedded in the slot layer is around 7.8 (notice that the
$\Gamma^{\rm TM}$ emission rate is dominated by $\Gamma_{\perp}^{\rm TM}$,
as the TM electric field component in the $xy$ plane has a very small amplitude in the slot
layer). Thus, the agreement between the theoretical results for the slot
waveguide obtained within our model and the measurements of Ref.~\cite{Galli06}
is quite satisfactory.

\section{Conclusions}
We have presented a quantum electrodynamical formalism in
order to analyze spontaneous emission in generic lossless and non-dispersive
multilayer dielectric structures. A second quantized form for the electromagnetic field,
which follows after its expansion into normal modes, has been set up and
used to derive the local density of states and express the decay rate $\Gamma$
as a function of the excited dipole position in the considered structure.
The expressions derived have been used to study the spontaneous emission in Si-based waveguides of
different geometries and with realistic parameters. The following conclusions summarize our results.

(i) The standard basis of radiative states generally used in the
description of the electromagnetic field modes in a dielectric
structure, based on incident/reflected/transmitted waves, is not the
most appropriate one in radiation emission analysis as it leads to a subtle interference 
between different outgoing components~\cite{Zakowicz95a,Glauber95}. 
By choosing a set of modes specified by a single outgoing radiative component, 
the total emission rate as well as the emission in the upper/lower claddings (more generally, 
the SE patterns) can be calculated in a simple way, without any interference issue. 
One basis can be transformed into
the other after application of the Time-Reversal operation.

(ii) The evanescent component of partially radiative modes which
arise in any asymmetric configuration, i.e. when the upper and lower
claddings have different refractive indices, is relevant for the SE
analysis, as it can be seen in the single interface as well as in
more complicated asymmetric structures.

(iii) We have calculated and compared SE rates for symmetric
(air/Si/air) and asymmetric (SiO$_2$/Si/air) silicon waveguides and
shown that, in the latter configuration, the lower index contrast
leads to an increased emission into radiative modes. Such an effect
is much more evident in a silicon Slot waveguide: in this
configuration the discontinuity of the normal component of the
electromagnetic field which develops at the high-index-contrast
interfaces of the slot layer, results into an enhancement of the
local density of states for TM polarized guided modes. As an
example, we have analyzed the SE rate in a Si Slot waveguide with
the same structure parameters used in Ref.~\cite{Galli06} and found
a very good agreement with the experimental evidence of the
enhancement of the TM/TE photoluminescence. Thus, the model
developed turns out to be a useful tool for the analysis of
spontaneous emission processes in realistic structures such as SOI
Slot waveguides. Further work will focus on analyzing more complex
slot waveguides, as well as photonic crystal slab structures.

\section*{Acknowledgments}
This work has been partially supported by the Piedmont Regional
Project ``Nanostructures for applied photonics (2004)'' and by Fondazione CARIPLO. The authors are grateful to
Fabrizio Giorgis (Politecnico di Torino) for encouragement and
support and to Dario Gerace (Universit$\rm{\grave{a}}$ degli Studi
di Pavia) for carefully reading the manuscript.

\appendix
\section{\label{subapp:rad_modes}Radiative modes}
With reference to the geometry of Fig.~\ref{fig:structure}, the
field amplitudes of the TE-polarized radiative modes [see
Eqs.~(\ref{eq:TE-rad-profs-E0}) and (\ref{eq:TE-rad-profs-H0}) in
Sec.~\ref{subsec:rad_modes}] are given by
\begin{widetext}
\begin{equation}\label{eq:TE-rad-profs-E1}
    \displaystyle
    E^{\rm TE}(\kp,z)=
    \left\{\begin{array}{ll}
    W_{\rm M+1}e^{iq_{\rm M+1}(z - z_{\rm M+1})} +
    X_{\rm M+1}e^{-iq_{\rm M+1}(z - z_{\rm M+1})}\,, & z> z_{\rm M+1}\\
    \displaystyle
    W_{j}e^{i q_{j}(z- z_{j} - d_{j}/2)} + X_{j}e^{-i q_{j}(z- z_{j} - d_{j}/2)}\,, & z_{j} < z < z_{j} + d_{j}=z_{j+1}\\
    W_{0}e^{iq_{0}(z - z_1)} + X_{0}e^{-iq_{0}(z - z_1)}\,, & z < z_1
    \end{array}\right.
\end{equation}
\begin{equation}\label{eq:TE-rad-profs-H1}
    \displaystyle
    H^{\rm TE}_{\perp}(\kp,z)=
    \left\{\begin{array}{ll}
    \kp[W_{\rm M+1}e^{iq_{\rm M+1}(z - z_{\rm
    M+1})}+ X_{\rm M+1}e^{-iq_{\rm M+1}(z - z_{\rm M+1})}]\,, & z> z_{\rm M+1}\\
    \displaystyle
    \kp[W_{j}e^{i q_{j}(z- z_{j} - d_{j}/2)}
    + X_{j}e^{-i q_{j}(z- z_{j} - d_{j}/2)}]\,, & z_{j} < z < z_{j} + d_{j}=z_{j+1}\\
    \kp[W_{0}e^{iq_{0}(z - z_1)} + X_{0}e^{-iq_{0}(z - z_1)}]\,, & z < z_1
    \end{array}\right.
\end{equation}
\begin{equation}\label{eq:TE-rad-profs-H2}
    \displaystyle
    H^{\rm TE}_{\parallel}(\kp,z)=
    \left\{\begin{array}{ll}
     q_{\rm M+1}[X_{\rm M+1}e^{-iq_{\rm M+1}(z - z_{\rm M+1})} - W_{\rm M+1}e^{iq_{\rm M+1}(z - z_{\rm
    M+1})}]\,, & z> z_{\rm M+1}\\
    \displaystyle
    q_{j}[X_{j}e^{-i q_{j}(z- z_{j} - d_{j}/2)} - W_{j}e^{i q_{j}(z- z_{j} -
    d_{j}/2)}]\,, & z_{j} < z < z_{j} + d_{j}=z_{j+1}\\
    q_{0}[X_{0}e^{-iq_{0}(z - z_1)} - W_{0}e^{iq_{0}(z - z_1)}]\,. & z < z_1
    \end{array}\right.
\end{equation}
\end{widetext}
For fully radiative modes outgoing in the lower (upper) cladding
[see Fig.~\ref{fig:rad-modes}(b)], $\rm W_{M+1}=0$ ($\rm X_{0}=0$)
in Eqs.~(\ref{eq:TE-rad-profs-E1})-(\ref{eq:TE-rad-profs-H2}) and
the amplitude $\rm X_0$ ($\rm W_{M+1}$) obtained through the
normalization condition Eq.~(\ref{eq:ortho-H}) is given by $\rm
X_{0}=1/\sqrt{\eps_0}$ ($\rm W_{M+1}=1/\sqrt{\eps_{\rm M+1}}$); all
the other coefficients are then found by application of standard
transfer-matrix theory. These results can be formally obtained by  
taking into account a normalization box having width $L$ in the $z$-direction: 
when $L\gg\,d$, $d$ being the thickness of the waveguide core 
or the thickness of a stack of layers in a generic multilayer structure, the 
contributions from the core/stack are of the order $O(d/L)$ and are 
negligibly small as compared to the contributions from the semi-infinite cladding regions. 
Thus, the normalization of the radiative modes is determined by the cladding regions only, 
and the values given above are found for the amplitudes $\rm
X_{0}$ and  $\rm W_{M+1}$. When the dielectric constants of the upper
and lower claddings are different and the conditions for total
internal reflection are matched, the modes become partially
radiative. Without loss of generality, we assume
$\eps_{0}>\eps_{\rm M+1}$. In this case, when $\displaystyle
\w\frac{\sqrt{\eps_{\rm M+1}}}{c} \leq \kp \leq
\w\frac{\sqrt{\eps_{\rm 0}}}{c}$, the emission occurs in the lower
cladding only and the field becomes evanescent in the upper
cladding, the $z$ component $q_{\rm M+1}$ being purely imaginary.
The field amplitudes are then found through the same conditions
given above for the fully radiative modes, together with the
transformation rule Eq.~(\ref{eq:TR-rule}), i.e., by taking $\rm
W_{M+1}=0$ and replacing $q_{\rm M+1}$ with its complex conjugate in
Eqs.~(\ref{eq:TE-rad-profs-E1})-(\ref{eq:TE-rad-profs-H2}).

For TM-polarized radiative modes [see Eqs.~(\ref{eq:TM-rad-profsH0})
and (\ref{eq:TM-rad-profsE0}) in Sec.~\ref{subsec:rad_modes}] the
field amplitudes  are given by:
\begin{widetext}
\begin{equation}\label{eq:TM-rad-profs-H1}
    \displaystyle
    H^{\rm TM}(\kp,z)=
    \left\{\begin{array}{ll}
    Y_{\rm M+1}e^{iq_{\rm M+1}(z - z_{\rm M+1})} +
    Z_{\rm M+1}e^{-iq_{\rm M+1}(z - z_{\rm M+1})}\,, & z> z_{\rm M+1}\\
    \displaystyle
    Y_{j}e^{i q_{j}(z- z_{j} - d_{j}/2)} + Z_{j}e^{-i q_{j}(z- z_{j} - d_{j}/2)}\,, & z_{j} < z < z_{j} + d_{j}=z_{j+1}\\
    Y_{0}e^{iq_{0}(z - z_1)} + Z_{0}e^{-iq_{0}(z - z_1)}\,, & z < z_1
    \end{array}\right.
\end{equation}
\begin{equation}\label{eq:TM-rad-profs-E1}
    \displaystyle
    E_{\perp}^{\rm TM}(\kp,z)=
    \left\{\begin{array}{ll}
    \kp[Y_{\rm M+1}e^{iq_{\rm M+1}(z - z_{\rm
    M+1})}+ Z_{\rm M+1}e^{-iq_{\rm M+1}(z - z_{\rm M+1})}]\,, & z> z_{\rm M+1}\\
    \displaystyle
    \kp[Y_{j}e^{i q_{j}(z- z_{j} - d_{j}/2)} + Z_{j}e^{-i q_{j}(z- z_{j} - d_{j}/2)}]
    \,, & z_{j} < z < z_{j} + d_{j}=z_{j+1}\\
    \kp[Y_{0}e^{iq_{0}(z - z_1)} + Z_{0}e^{-iq_{0}(z - z_1)}\,, & z < z_1
    \end{array}\right.
\end{equation}
\begin{equation}\label{eq:TM-rad-profs-E2}
    \displaystyle
    E_{\parallel}^{\rm TM}(\kp,z)=
    \left\{\begin{array}{ll}
    q_{\rm M+1}[Z_{\rm M+1}e^{-iq_{\rm M+1}(z - z_{\rm M+1})}-Y_{\rm M+1}e^{iq_{\rm M+1}(z - z_{\rm
    M+1})}]\,, & z> z_{\rm M+1}\\
    \displaystyle
    q_{j}[Z_{j}e^{-i q_{j}(z- z_{j} - d_{j}/2)}-Y_{j}e^{i q_{j}(z- z_{j} - d_{j}/2)}
    ]\,, & z_{j} < z < z_{j} + d_{j}=z_{j+1}\\
    q_{0}[Z_{0}e^{-iq_{0}(z - z_1)}-Y_{0}e^{iq_{0}(z - z_1)}]\,. & z < z_1
    \end{array}\right.
\end{equation}
\end{widetext}
Notice that $E^{\rm TM}$ defined in Eqs.~(\ref{eq:TM-rad-profs-E1}),
(\ref{eq:TM-rad-profs-E2}) have the same dimensions of $H^{\rm TE}$
defined in Eqs.~(\ref{eq:TE-rad-profs-H1}),
(\ref{eq:TE-rad-profs-H2}), while $H^{\rm TM}$ defined in
Eq.~(\ref{eq:TM-rad-profs-H1}) has the same dimensions of $E^{\rm
TM}$ defined in Eq.~(\ref{eq:TE-rad-profs-E1}). For fully radiative
modes outgoing in the lower (upper) cladding, $\rm Y_{M+1}=0$ ($\rm
Z_{0}=0$) and the normalization condition Eq.~(\ref{eq:ortho-H})
yields $\rm Z_{0}=1$ ($\rm Y_{M+1}=1$) for the amplitude of the
outgoing component. As for TE-polarized modes, all the other
coefficients are straightforwardly found after a standard
transfer-matrix calculation. For modes which are partially radiative
in the lower cladding (evanescent in the upper cladding), one takes
$\rm Y_{\rm M+1}=0$ and replaces $q_{\rm M+1}$ with its complex
conjugate in
Eqs.~(\ref{eq:TM-rad-profs-H1})-(\ref{eq:TM-rad-profs-E2}).

\section{\label{subapp:gui_modes}Guided modes}
The field amplitudes for TE-polarized guided modes [see
Eqs.~(\ref{eq:TE-guided-profs-E0}) and (\ref{eq:TE-guided-profs-H0})
in Sec.~\ref{subsec:rad_modes}]
\begin{widetext}
\begin{equation}\label{eq:TE-guided-profs-E1}
    E^{\rm TE}(\kp,z)=
    \left\{\begin{array}{ll}
    A_{\rm M+1\,\mu}e^{-\chi_{\rm M+1\,,\mu}(z - z_{\rm M+1})}\,, & z> z_{\rm M+1}\\
    \displaystyle
    A_{j\,\mu}e^{i q_{j\,\mu}(z- z_{j} - d_{j}/2)} + B_{j\,\mu}e^{-i q_{j\,\mu}(z- z_{j} - d_{j}/2)}\,, & z_{j} < z < z_{j} + d_{j}=z_{j+1}\\
    B_{0\,\mu}e^{\chi_{0\,\mu}(z - z_1)}\,, & z < z_1
    \end{array}\right.
\end{equation}
\begin{equation}\label{eq:TE-guided-profs-H1}
    H_{\perp}^{\rm TE}(\kp,z)=
    \left\{\begin{array}{ll}
    iA_{\rm M+1\,\mu}\kp e^{-\chi_{\rm M+1\,\mu}(z - z_{\rm M+1})}\,, & z> z_{\rm M+1}\\
    i\kp[A_{j\,\mu}e^{i q_{j\,\mu}(z- z_{j} -
    d_{j}/2)} + B_{j\,\mu}e^{-i q_{j\,\mu}(z- z_{j} - d_{j}/2)}]\,, & z_{j} < z < z_{j} + d_{j}=z_{j+1}\\
    iB_{0\,\mu}\kp\hat{z})e^{\chi_{0\,\mu}(z - z_1)}\,, & z < z_1
    \end{array}\right.
\end{equation}
\begin{equation}\label{eq:TE-guided-profs-H2}
    H_{\parallel}^{\rm TE}(\kp,z)=
    \left\{\begin{array}{ll}
    A_{\rm M+1\,\mu}\chi_{\rm M+1\,\mu}e^{-\chi_{\rm M+1\,\mu}(z - z_{\rm M+1})}\,, & z> z_{\rm M+1}\\
    iq_{j\,\mu}[B_{j\,\mu}e^{-i q_{j\,\mu}(z- z_{j} - d_{j}/2)}-A_{j\,\mu}e^{i q_{j\,\mu}(z- z_{j} - d_{j}/2)}]\,, & z_{j} < z < z_{j} + d_{j}
    =z_{j+1}\\
    -B_{0\,\mu}\chi_{0\,\mu}e^{\chi_{0\,\mu}(z - z_1)}\,, & z < z_1
    \end{array}\right.
\end{equation}
\end{widetext}
where $S$ is a normalization surface which cancels in the final
expressions for the emission rates, and the magnetic field is found
by application of the Maxwell equation $\displaystyle
\mathbf{H}(\mathbf{r})=-\frac{i\,c}{\w}\mathbf{\nabla}\times\mathbf{E}(\mathbf{r})$.
The M+2 coefficients in the expressions
Eqs.~(\ref{eq:TE-guided-profs-E1}), (\ref{eq:TE-guided-profs-H1})
and (\ref{eq:TE-guided-profs-H2}) are found by solving the system
consisting of M+1 relations which follow from the application of
standard transfer-matrix theory and the orthormality condition
Eq.~(\ref{eq:ortho-H}) which leads to:
\begin{widetext}
\begin{eqnarray}\label{eq:ortho_cond_explicit-TE}
\displaystyle
    \int\mid\mathbf{H}(\rhob,z)\mid^{2}\ud\rhob\ud
    z&=&1=\frac{\chi_{0}^2 + \kp^2}{2\chi_0}|B_0|^2 +
    \frac{\chi_{\rm M+1}^2 + \kp^2}{2\chi_{\rm M+1}}|A_{\rm
    M+1}|^2\nonumber\\
    &+& \sum_{j=1}^{\rm M}d_{j}\left[\left(\kp^2 + q_{j}q_{j}^{*}\right)
    \left(|A_j|^2 + |B_j|^2\right)\textrm{sinc}\left(\frac{\left(q_j -
    q_{j}^{*}\right)d_j}{2}\right)\right.\nonumber\\
    &+&\left.\left(\kp^2 - q_{j}q_{j}^{*}\right)
    \left(A_{j}^{*}B_{j} + B_{j}^{*}A_{j}\right)\textrm{sinc}\left(\frac{\left(q_j
    + q_{j}^{*}\right)d_j}{2}\right)\right]\,,
\end{eqnarray}
\end{widetext}
with $\rm sinc(x)=sin(x)/x$.  For TM-polarized guided modes [see
Eqs.~(\ref{eq:TM-guided-profs-H0}) and (\ref{eq:TM-guided-profs-E0})
in Sec.~\ref{subsec:rad_modes}] the field amplitudes are given by
\begin{widetext}
\begin{equation}\label{eq:TM-guided-profs-H1}
    H^{\rm TM}(\kp,z)=
    \left\{\begin{array}{ll}
    C_{\rm M+1\,\mu}e^{-\chi_{\rm M+1\,,\mu}(z - z_{\rm M+1})}\,, & z> z_{\rm M+2}\\
    C_{j\,\mu}e^{i q_{j\,\mu}(z- z_{j} - d_{j}/2)} + D_{j\,\mu}e^{-i q_{j\,\mu}(z- z_{j} - d_{j}/2)}\,,
     & z_{j} < z < z_{j} + d_{j}=z_{j+1}\\
    D_{0\,\mu}e^{\chi_{0\,\mu}(z - z_1)}\,, & z < z_1
    \end{array}\right.
\end{equation}
\begin{equation}\label{eq:TM-guided-profs-E1}
\displaystyle
    E^{\rm TM}_{\perp}(\kp,z)=
    \left\{\begin{array}{ll}
    \displaystyle
    \frac{i}{\eps_{\rm M+1}}C_{\rm M+1\,\mu}\kp e^{-\chi_{\rm M+1\,\mu}(z - z_{\rm M+1})}\,, & z> z_{\rm M+1}\\
    \displaystyle
    \frac{i}{\eps_j}\kp[C_{j\,\mu}e^{i q_{j\,\mu}(z- z_{j} - d_{j}/2)} + D_{j\,\mu}
    e^{-i q_{j\,\mu}(z- z_{j} - d_{j}/2)}]\,, & z_{j} < z < z_{j} + d_{j}=z_{j+1}\\
    \displaystyle
    \frac{i}{\eps_1}D_{0\,\mu}\kp e^{\chi_{0\,\mu}(z - z_1)}\,, & z < z_1
    \end{array}\right.
\end{equation}
\begin{equation}\label{eq:TM-guided-profs-E2}
\displaystyle
    E^{\rm TM}_{\parallel}(\kp,z)=
    \left\{\begin{array}{ll}
    \displaystyle
    \frac{1}{\eps_{\rm M+1}}C_{\rm M+1\,\mu}\chi_{\rm M+1\,\mu}e^{-\chi_{\rm M+1\,\mu}
    (z - z_{\rm M+1})}\,, & z> z_{\rm M+1}\\
    \displaystyle
    \frac{i}{\eps_j}q_{j\,\mu}[D_{j\,\mu}e^{-i q_{j\,\mu}
    (z- z_{j} - d_{j}/2)}-C_{j\,\mu}e^{i q_{j\,\mu}
    (z- z_{j} - d_{j}/2)}]\,, & z_{j} < z < z_{j} + d_{j}=z_{j+1}\\
    \displaystyle
    -\frac{1}{\eps_0}D_{0\,\mu}\chi_{0\,\mu}e^{\chi_{0\,\mu}(z - z_1)}\,, & z < z_1
    \end{array}\right.
\end{equation}
\end{widetext}
where the electric field is obtained from the relation
$\displaystyle
\mathbf{E}(\mathbf{r})=\frac{ic}{\w\eps(\mathbf{r})}\mathbf{\nabla}\times\mathbf{H}(\mathbf{r})$.
As for TE-polarized modes, the M+2 coefficients in the above
expressions are derived within the transfer-matrix theory together
with normalization integral Eq.~(\ref{eq:ortho-H}) which yields the
condition:
\begin{widetext}
\begin{eqnarray}\label{eq:ortho_cond_explicit-TM}
\displaystyle
    \int\mid\mathbf{H}(\rhob,z)\mid^{2}\ud\rhob\ud
    z&=&1=\frac{|D_0|^2}{2\chi_0} +
    \frac{|C_{\rm M+1}|^2}{2\chi_{\rm M+1}}|\nonumber\\
    &+& \sum_{j=1}^{\rm M}d_{j}\left[\left(|C_j|^2 + |D_j|^2\right)\textrm{sinc}\left(\frac{\left(q_j -
    q_{j}^{*}\right)d_j}{2}\right)\right.\nonumber\\
    &+&\left.\left(C_{j}^{*}D_{j} +
D_{j}^{*}C_{j}\right)\textrm{sinc}\left(\frac{\left(q_j
    + q_{j}^{*}\right)d_j}{2}\right)\right]\,.
\end{eqnarray}
\end{widetext}


\begin{thebibliography}{50}
\bibitem[{not({\natexlab{a}})}]{note_radiative}
\bibinfo{note}{The word ``radiative'' is here used to identify a specific class
  of modes supported by a dielectric waveguide as commonly done in the
  literature, and does not embody all the spatial modes which contribute to the
  total radiative spontaneous emission. In the literature, ``radiative'' modes
  are also known as ``traveling'' modes in contrast with ``guided'' modes which
  are sometimes called ``trapped'' modes (Ref. 18).}
\bibitem[{not({\natexlab{b}})}]{note_gamma}
\bibinfo{note}{Furthermore, it can be proved that, for a generic interface made
  by two dielectric half-spaces with index $n_0$ (at $z<0$) and $n_1$ (at
  $z>0$) and in the limit $n_{0}\gg\,n_{1}$, $\Gamma(z\rightarrow
  0^-)/\Gamma_{0}=n_{1}^{4}/n_{0}^{3}$ and $\Gamma(z\rightarrow
  0^+)/\Gamma_{0}=n_{0}$}.
\end{thebibliography}

\end{document}